\documentclass[11pt]{article}
\pdfoutput=1
\usepackage{putex}
\usepackage[vcentermath]{youngtab}
\usepackage{subfig}
\usepackage{lscape}

\usepackage{graphicx}
\usepackage{epstopdf}
\usepackage{enumerate}
\usepackage{cite}
\usepackage{tensor}
\usepackage{slashed}
\usepackage{amsmath}
\usepackage{amssymb}
\usepackage{mathrsfs}
\usepackage{lgrind}

\usepackage{bbm}

\usepackage{hyperref}

\numberwithin{equation}{section}

\newcommand {\be} {\begin {equation}}
\newcommand {\ee} {\end {equation}}

\newcommand {\bes} {\begin {equation*}}
\newcommand {\ees} {\end {equation*}}

\newcommand{\eps}{\epsilon}

\def\CH{{\cal H}}

\newcommand{\beq}{\begin{equation}}
\newcommand{\eeq}{\end{equation}}

\def\be{ \begin{equation} }
\def\ee{ \end{equation} }

\def\XXint#1#2#3{{\setbox0=\hbox{$#1{#2#3}{\int}$}
\vcenter{\hbox{$#2#3$}}\kern-.5\wd0}}

\def\tr{{\textrm{tr}}}

\def\???th{{\textrm{\,???th}}}
\def \eps {\epsilon}

\def\XXint#1#2#3{{\setbox0=\hbox{$#1{#2#3}{\int}$}
     \vcenter{\hbox{$#2#3$}}\kern-.5\wd0}}

\newcommand{\ignore}[1]{}

\begin{document}

\preprint{PUPT-2484}

\institution{PU}{Department of Physics, Princeton University, Princeton, NJ 08544}
\institution{PCTS}{Princeton Center for Theoretical Science, Princeton University, Princeton, NJ 08544}

\title{Conformal QED$_d$, $F$-Theorem and the $\epsilon$ Expansion}

\authors{Simone Giombi,\worksat{\PU} Igor R.~Klebanov\worksat{\PU,\PCTS} and Grigory Tarnopolsky\worksat{\PU}}

\abstract{
We calculate the free energies $F$ for $U(1)$ gauge theories on the $d$ dimensional sphere of radius $R$.
For the theory with free Maxwell action we find the exact result as a function of $d$;
it contains the term
$\frac{d-4}{2} \log R$ consistent with the lack of conformal invariance in dimensions other than 4.
When the $U(1)$ gauge theory is coupled to a sufficient number $N_f$ of massless 4-component fermions, it acquires an interacting conformal phase,
which in $d<4$ describes the long distance behavior of the model.
The conformal phase can be studied using large $N_f$ methods. Generalizing the $d=3$ calculation in
 arXiv:1112.5342, we compute its sphere free energy as a function of $d$,
ignoring the terms of order $1/N_f$ and higher.
For finite $N_f$, following arXiv:1409.1937 and arXiv:1507.01960, we develop the $4-\epsilon$ expansion for the sphere free energy of conformal QED$_d$. Its extrapolation to $d=3$
shows very good agreement with the large $N_f$ approximation for $N_f>3$.
For $N_f$ at or below some critical value $N_{\rm crit}$, the $SU(2N_f)$ symmetric conformal phase of QED$_3$ is expected to disappear or become unstable. By using the $F$-theorem and comparing the sphere free energies in the conformal and broken symmetry phases, we show that
 $N_{\rm crit}\leq 4$. As another application of our results, we
calculate the one loop beta function in conformal QED$_6$, where the gauge field has a 4-derivative kinetic term.
We show that this theory coupled to $N_f$ massless fermions is asymptotically free.
}

\date{}
\maketitle

\tableofcontents

\section{Introduction and Summary}

The four-dimensional Quantum Electrodynamics coupled to $N_f$ Dirac fermions is an original
model of Quantum Field Theory; its predictions have been verified experimentally with high accuracy. If the fermions are massless, then the theory is conformally invariant for zero charge $e$, but the interaction effects are well known to break the conformal invariance. They produce
a positive $\beta$ function for $e$, which means that the theory becomes free at long distances.

The physics of QED is different in $d\neq 4$. Then
the free Maxwell action ${1\over 4} F_{\mu\nu} F^{\mu\nu}$ is not conformally invariant \cite{ElShowk:2011gz}, but the one loop fermion vacuum polarization diagram induces a scale invariant quadratic term proportional to
\begin{equation}
F_{\mu \nu} (-\nabla^2)^{\frac{d}{2}-2} F^{\mu \nu}\ ,
\label{induced}
\end{equation}
which is in general non-local. For $d<4$ this term dominates at long distances, and well-known examples of such ``induced QED'' are the Schwinger model \cite{Schwinger:1962tp} in $d=2$ and the conformal phase of QED$_3$
\cite{Appelquist:1981vg,Appelquist:1988sr}.
In $d=4-\eps$ the conformal QED$_d$ theory may be studied using
the $\epsilon$ expansion, because the $\beta$ function
\begin{align}
\beta=-\frac{\epsilon }{2}e+\frac{4N_{f}}{3}\frac{e^3}{ (4 \pi )^2}+{\cal O} (e^5)
\label{beta-QED-oneloop}
\end{align}
has a weakly coupled IR fixed point at $e_*^2=6\epsilon \pi^2/N_f + {\cal O} (\epsilon^2)$
\cite{Moshe:2003xn}. The $\epsilon$ expansion of various operator dimensions in
QED$_d$ was introduced in \cite{Beneke:1995qq,Zohar}.

For $d>4$ the induced term (\ref{induced}) becomes important at short distances (for $d=6,8,\ldots$
it is a local higher derivative term). A perturbatively renormalizable theory in $d=6$ based
on the local 4-derivative action ${1\over 4}F_{\mu \nu} (-\nabla^2) F^{\mu \nu}$ was studied in \cite{Ivanov:2005qf,Ivanov:2005kz,Smilga:2005pr,Smilga:2006ax} and, more recently, in \cite{Beccaria:2015uta}.\footnote{Its Weyl anomaly coefficient $a=-55/84$ may be calculated either using a dual AdS approach \cite{Giombi:2013yva,Giombi:2014iua} or directly in $d=6$ \cite{Tseytlin:2013fca}. }
In section \ref{comments} we will calculate the one loop beta function for this 6-d theory coupled to $N_f$ Weyl fermions; it has the negative sign which produces the asymptotic freedom.

Among the important physical applications of the conformal QED is the theory in $d=3$ coupled to massless Dirac fermions and/or
complex scalars. An early motivation to study QED$_3$ came from work on the high temperature behavior of four-dimensional gauge theory \cite{Appelquist:1981vg}. More recently, its various applications to condensed matter physics have been explored as well (see, for example, \cite{Marston:1989zz,Franz:2002qy,Herbut:2002yq}). Work on QED$_3$ has uncovered a variety of interesting phenomena, which include
chiral symmetry breaking and interacting conformal field theory \cite{Pisarski:1984dj,Appelquist:1986fd,Appelquist:1988sr}. Both of these phases of the theory are consistent with the Vafa-Witten theorem \cite{Vafa:1984xh}, which requires the presence of massless modes for $N_f>3$. Yet, some questions remain about the infrared behavior of the theory. In this paper we re-examine them using the relatively new tools provided by the $F$-theorem \cite{Jafferis:2011zi,Klebanov:2011gs,Myers:2010xs,Casini:2011kv}. Our analysis is similar in spirit to that of \cite{Grover:2012sp,Dyer:2013fja}, although some of our reasoning is different.

Specifically, we will work with the $U(1)$ gauge theory coupled to $N_f$ massless 4-component Dirac fermions $\psi^j$. The lagrangian of this theory has $SU(2N_f)$ global symmetry, which is often referred to as the ``chiral symmetry.'' In QED$_3$ the fine structure constant $\alpha={e^2\over 4\pi}$ has dimension of mass; this makes the theory super-renormalizable. At short distances we find a weakly interacting theory of massless fermions and photons, where the field strength $F_{\mu\nu}$ has scaling dimension $3/2$. The short distance limit of QED$_3$ is scale invariant, but not conformal. This is because the free Maxwell action ${1\over 4} F_{\mu\nu} F^{\mu\nu}$ is not conformally invariant in three dimensions \cite{ElShowk:2011gz}. The lack of conformal invariance of the free Maxwell theory translates into the fact that its three-sphere free energy $F$ depends logarithmically on the sphere radius $R$
\cite{Klebanov:2011td}. In section \ref{freeMaxwell} we will generalize this result
to free Maxwell theory on $S^d$ and show that its free energy contains the term
$\frac{d-4}{2} \log R$.
We will refer to the short distance limit of QED$_3$ as the UV theory. The fact that its $F$ value, $F_{\rm UV}$, diverges is important for consistency of the RG flows with the $F$-theorem.

As QED$_3$ flows to longer distances, the effective interaction strength grows and various
interesting phenomena become possible. The one loop fermion vacuum polarization diagram induces a non-local quadratic term (\ref{induced}) for $A_\mu$, which dominates in the IR over the Maxwell term \cite{Appelquist:1981vg}. Due to this effect, the theory flows to an interacting conformal field theory in the large $N_f$ limit where $e^2 N_f$ is held fixed. In the CFT the
scaling dimension of $F_{\mu \nu}$ is $2$. The scaling dimensions of other operators can be calculated as series in $1/N_f$ (see, for example, \cite{Gracey:1993iu,PhysRevB.77.155105,2008PhRvB..78e4432X}).

A different possibility is the spontaneous breaking of the $SU(2N_f)$ global symmetry due to generation of vacuum expectation value of the operator $\sum_{j=1}^{N_f} \bar \psi_j \psi^j$ (it is written using the 4-d notation for spinors $\psi^i$ and gamma-matrices). This
operator preserves the 3-d parity and time reversal symmetries,
but it breaks the global symmetry to $SU(N_f)\times SU(N_f)\times U(1)$.
This mechanism was proposed in \cite{Pisarski:1984dj}, where it was argued using Schwinger-Dyson equations to be possible for any $N_f$;
however, for large $N_f$ the scale of the VEV becomes exponentially small compared to $\alpha$.
Subsequently, modified treatments of the Schwinger-Dyson equations \cite{Appelquist:1988sr} suggested that the chiral symmetry breaking is possible only for $N_f \leq N_{\rm crit}$.
The estimates of $N_{\rm crit}$ typically range between $2$ and $10$ \cite{Kaveh:2004qa,Fischer:2004nq,Braun:2014wja, Zohar}.

It is widely believed that the QED$_3$ must be in the conformal phase for
$N_f > N_{\textrm{crit}}$, but
 a nearly marginal operator may appear in the spectrum of the CFT as $N_f$ is reduced towards $N_{\textrm{crit}}$.
This operator must respect the
$SU(2N_f)$ and parity symmetries of the theory, and natural candidates are the operators quartic in the fermion fields\footnote{We are grateful to Z. Komargodski for informing us of this possibility.} \cite{Zohar}  (see also \cite{2008PhRvB..78e4432X,Braun:2014wja}).\footnote{
In the compact theory, monopole operators may also become relevant as one lowers $N_f$ \cite{Pufu:2013vpa}; however, these operators 
transform in non-trivial representations of the $SU(2N_f)$ flavor symmetry, and so they are not expected to be generated along the RG flow if the 
UV theory has exact $SU(2N_f)$ symmetry. Monopoles may still condense, i.e. they may acquire expectation values in the spontaneously 
broken phase.}
When the quartic operator is slightly irrelevant, it should give rise to a nearby UV fixed point; there is a standard argument for this using conformal perturbation theory, which we present in section \ref{PadeSec}. We will call this additional fixed point QED$_3^*$.
 For $N_f = N_{\textrm{crit}}$ it merges with QED$_3$, and for
 $N_f < N_{\textrm{crit}}$ both fixed points may become complex \cite{Kaveh:2004qa,Gies:2005as,Kaplan:2009kr,Braun:2014wja}.
 In this ``merger and annihilation of fixed points" scenario, for $N_f < N_{\textrm{crit}}$ the UV theory flows directly to the broken symmetry phase.
   Alternatively, both fixed points may stay real and go through each other. Then the QED$_3$ fixed point continues to exist even after the appearance of a relevant operator; this relevant operator may create flow from QED$_3$ to the broken symmetry phase. If so, the edge of the conformal window may be associated with the dimension of some operator in QED$_3$ becoming so small that it violates the unitarity bound. This would be analogous to what happens at the lower edge of the conformal window in the
${\cal N}=1$ supersymmetric gauge theory \cite{Seiberg:1994pq}.

We will attempt to shed new light on the transition from the conformal to the symmetry breaking behavior by using the $F$-theorem and performing more precise calculations of $F$. Here $F=-\log Z_{S^3}$ is the 3-sphere free energy \cite{Jafferis:2011zi,Klebanov:2011gs} or, equivalently, the long-range Entanglement Entropy across a circle \cite{Myers:2010xs,Casini:2011kv}. The theorem states that for Renormalization Group (RG) flow from fixed point 1 to fixed point 2,
$F_1> F_2$. A proof of this inequality has been found using properties of the Renormalized Entanglement Entropy in relativistic field theories \cite{Casini:2012ei} (see also \cite{Liu:2012ee}).

In order to apply the $F$-theorem to RG flows among different phases of QED$_3$, it is important to know their $F$-values. This is especially challenging for the interacting CFT phase of the theory.
In \cite{Klebanov:2011td} this calculation was performed using the $1/N_f$ expansion with the result
\begin{equation}
\label{confQED}
F_{\rm conf} = N_f \left(\frac{\log(2)}{2}+\frac{3\zeta(3)}{4\pi^2}\right)+\frac{1}{2}\log\left(\frac{\pi N_f}{4}\right)+{\cal O}(\frac{1}{N_f})\,.
\end{equation}
The first term on the RHS is the $F$-value of $N_f$ free Dirac fermions, $N_{f}F_{\rm free-ferm}$.
Even though $F_{\rm conf}-N_{f}F_{\rm free-ferm}$ grows without bound for large $N_f$, the $F$-theorem inequality
$F_{\rm UV}> F_{\rm conf}$ is satisfied. This is because $F_{\rm UV}$ is infinite due to the divergent
contribution of the free Maxwell theory. In section \ref{largeN} we review the large $N$ description of conformal QED and generalize the result (\ref{confQED}) by computing $F_{\rm conf}$ as a function of $d$.

Since we will be quite interested in $F_{\rm conf}$ for small $N_f$, in this paper we will apply a different
approximation method \cite{Giombi:2014xxa,Fei:2015oha}. This method consists of developing the $\epsilon$ expansion of $\tilde F=-\sin (\pi d/2) F_{S^{d}}$ for $d=4-\eps$. It relies on the perturbative renormalization of the field theory on the sphere $S^{4-\epsilon}$ and requires inclusion of counter terms that involve the curvature tensor \cite{Drummond:1977dg,Brown:1980qq, Hathrell:1981zb,Jack:1983sk,Jack:1990eb}. Applications of this method to
the Wilson-Fisher $O(N)$ symmetric CFTs have produced high-quality estimates of $F_{\textrm{O(N)}}$ in $d=3$; they are found to be only $2-3\%$ below the $F$ values for the corresponding free UV fixed points of these theories \cite{Giombi:2014xxa,Fei:2015oha}.

In this paper, we will perform a similar $\epsilon$ expansion for $\tilde F$ of the conformal QED, building on earlier work which developed the perturbative renormalization of QED on $S^{4-\epsilon}$
\cite{Drummond:1977uy, Hathrell:1981gz, Jack:1990eb, Shore:1978hj, Harris:1993gd}. This calculation is presented in section \ref{F-eps}, and our main result is
\begin{equation}
\begin{aligned}
&\tilde{F}_{\rm conf}= N_{f}\tilde{F}_{\textrm{free-ferm}} -\frac{1}{2}\sin(\frac{\pi d}{2}) \log(\frac{N_{f}}{\epsilon}) \\
&~~~~~~~~~+
\frac{31\pi}{90}-1.2597 \eps-0.6493 \eps^2+0.8429 \eps^3+\frac{0.4418 \eps^2}{N_{f}}-\frac{0.6203 \eps^3}{N_{f}}-\frac{0.5522 \eps^3}{N_{f}^2}+ {\cal O}(\eps^4)
\,.
\label{tF-conf}
\end{aligned}
\end{equation}
Extrapolating this to $d=3$ using Pad\'e approximants produces results very close to the large $N_f$ formula
(\ref{confQED}) already for $N_f>3$, see figure \ref{PadevsLN}.

Applying the $F$-theorem, we find that RG flow from the conformal to the broken symmetry phase is impossible when $F_{\rm conf} < F_{\rm SB}$. This puts an upper bound on the value $N_{\rm crit}$ where the conformal phase can become unstable \cite{Grover:2012sp}. Using our resummed $\epsilon$ expansion results for $\tilde{F}_{\rm conf}$, we find that the value of $N_f$ where $F_{\rm conf} = F_{\rm SB}$ rather robustly lies between $4$ and $5$, and our best estimate is $N_f \approx 4.4$. If we restrict to integer values of $N_f$, this means that for $N_f\geq 5$ the QED$_3$ theory must be in the $SU(2N_f)$ symmetric conformal phase. Therefore, our results give the upper bound $N_{\rm crit}\leq 4$. The same upper bound is obtained if we use the large $N_f$ approximation (\ref{confQED})
to $F_{\rm conf}$, which was derived in \cite{Klebanov:2011td}.
The results obtained very recently using the $\epsilon$ expansion of quartic operator dimensions \cite{Zohar}, as well as computations in lattice gauge theory \cite{Strouthos:2008kc,Raviv:2014xna},  are consistent with our upper bound.

\section{Free energy of Maxwell theory on $S^d$}
\label{freeMaxwell}

The action for Maxwell theory on a curved manifold is
\begin{eqnarray}
S = \int d^d x \sqrt{g} \frac{1}{4e^2}F_{\mu\nu}F^{\mu\nu} =
\frac{1}{2e^2}\int d^d x \sqrt{g} A^{\nu}\left(-\delta^{\mu}_{\nu}\nabla^2+R^{\mu}_{\nu}+\nabla_{\nu}\nabla^{\mu}\right)A_{\mu} \,,
\end{eqnarray}
where we have used $F_{\mu\nu}=\nabla_{\mu}A_{\nu}-\nabla_{\nu}A_{\mu}$ and $[\nabla^{\mu},\nabla_{\nu}]A_{\mu} = R^{\mu}_{\nu}A_{\mu}$. On
a round $S^d$ of radius $R$, we have $R^{\mu}_{\nu}=\frac{d-1}{R^2}\delta^{\mu}_{\nu}$ and so the action is
\begin{equation}
S = \int_{S^d} d^d x \sqrt{g}\frac{1}{2e^2}  A^{\nu}\left(\delta^{\mu}_{\nu}(-\nabla^2+\frac{d-1}{R^2})+\nabla_{\nu}\nabla^{\mu}\right)A_{\mu}\,.
\end{equation}
The partition function is given by
\begin{equation}
Z = \frac{1}{{\rm vol}(G)} \int DA e^{-S(A)}\,,
\label{Zdef}
\end{equation}
where $G$ is the volume of the group of gauge transformations. One way to proceed is to split the gauge field into transverse and pure
gauge part\footnote{Equivalently, one can use Feynman gauge by adding a gauge fixing term $L_{\rm fix} =\frac{1}{2}(\nabla^{\mu}A_{\mu})^2$.
This gauge is more convenient for perturbative calculations when interactions with matter fields are included, and we will use it in Section \ref{F-eps}.}
\begin{equation}
A_{\mu} = B_{\mu}+\partial_{\mu}\phi\,,\qquad \nabla^{\mu}B_{\mu}=0\,.
\end{equation}
Following \cite{Klebanov:2011td}, we have
\begin{equation}
\begin{aligned}
&DA = DB D(d\phi) =DB D'\phi \sqrt{{\rm det}'(-\nabla^2)}\\
&{\rm vol}(G) = 2\pi \sqrt{{\rm vol}(S^d)}\int D'\phi\,,\qquad
{\rm vol}(S^d)=\frac{2\pi^{\frac{d+1}{2}}}{\Gamma\left(\frac{d+1}{2}\right)} R^d\equiv \Omega_d R^d\,,
\end{aligned}
\end{equation}
where prime means that the constant mode is not included. Then, the partition function can be written as
\begin{equation}
Z = \frac{\sqrt{{\rm det}'(-\nabla^2)}}{2\pi \sqrt{{\rm vol}(S^d)}}\int DB
e^{-\int_{S^d} d^d x \sqrt{g}\frac{1}{2e^2}  B^{\mu}\left(-\nabla^2+\frac{d-1}{R^2}\right)B_{\mu} }
= \frac{1}{2\pi \sqrt{{\rm vol}(S^d)}}\frac{\sqrt{{\rm det}'(-\nabla^2)}}{\sqrt{{\rm det}_{T}(\frac{-\nabla^2+(d-1)/R^2}{2\pi e^2})}}\,,
\label{ZMax}
\end{equation}
where the subscript `$T$' indicates that the determinant is taken on the space of transverse vector fields.

The eigenvalues of the sphere Laplacian $-\nabla^2$ acting on a transverse vector and corresponding degeneracies are known to be (see e.g.
\cite{Rubin:1984tc, Camporesi:1994ga})
\begin{equation}
\lambda_{\ell}^{(1)} = \frac{1}{R^2}(\ell (\ell+d-1)-1)\,,\qquad
g_{\ell}^{(1)} = \frac{\ell (\ell+d-1) (2 \ell+d-1) \Gamma \left(\ell+d-2\right)}{\Gamma\left(\ell+2\right) \Gamma\left(d-1\right)}\,,\quad \ell \ge 1\,.
\label{deg-1}
\end{equation}
For a scalar field, one has
\begin{equation}
\lambda_{\ell}^{(0)} = \frac{1}{R^2}\ell (\ell+d-1)\,,\qquad
g_{\ell}^{(0)} = \frac{(2 \ell+d-1) \Gamma \left(\ell+d-1\right)}{\Gamma \left(\ell+1\right) \Gamma \left(d\right)}\,,\quad \ell \ge 0\,.
\label{deg-0}
\end{equation}
In the case of the scalar field, $\ell=0$ corresponds to the constant mode which is to be excluded in our case. Using these results,
the free energy of Maxwell theory on $S^d$, $F_{\rm Maxwell} =-\log Z$, can be written as
\begin{equation}
F_{\rm Maxwell} = \frac{1}{2}\sum_{\ell=1}^{\infty} g_{\ell}^{(1)} \log(\frac{(\ell+1)(\ell+d-2)}{2\pi e^2 R^2})
-\frac{1}{2}\sum_{\ell=1}^{\infty} g_{\ell}^{(0)} \log(\frac{\ell(\ell+d-1)}{R^2})+\log(2\pi  \sqrt{{\rm vol}(S^d)})\,.
\label{FMaxwell}
\end{equation}
In dimensional regularization, the following results for the sum over vector and scalar degeneracies hold
\begin{equation}
\begin{aligned}
&\sum_{\ell=1}^{\infty} g_{\ell}^{(1)} = 1,\qquad \sum_{\ell=0}^{\infty} g_{\ell}^{(0)} = 0\,,\qquad \rightarrow \qquad \sum_{\ell=1}^{\infty} g_{\ell}^{(0)} = -1\,.
\label{deg-sum}
\end{aligned}
\end{equation}
These can be obtained for example by evaluating the sums for sufficiently negative $d$ where they converge, and analytically continuing
to positive values of $d$. Using these regularized identities, one can readily extract the radius dependence of the Maxwell free energy
(\ref{FMaxwell}) to be
\begin{equation}
F_{\rm Maxwell} = -\frac{1}{2}\log(e^2 R^{4-d})+F_{\rm Max.}^{(0)}(d)\,,
\label{FMax}
\end{equation}
where $F^{(0)}_{\rm Max.}(d)$ is a radius independent function of $d$ (with simple poles at even $d$). In particular, we see that $F_{\rm Maxwell} \rightarrow +\infty$ in the short
distance limit for $d<4$. The function $F^{(0)}_{\rm Max.}(d)$ can be evaluated in continuous $d$ by computing the non-trivial sums
in (\ref{FMaxwell}), as we explain below.

We first find it convenient to rewrite the free energy in the following way
\begin{equation}
F_{\rm Maxwell} = F_{\rm vector} -2 F_{\rm min-sc}+F_{\rm measure}\,,
\label{F-feyn}
\end{equation}
where we have defined
\begin{equation}
\begin{aligned}
&F_{\rm vector} = \frac{1}{2}\sum_{\ell=1}^{\infty} g_{\ell}^{(1)} \log(\frac{(\ell+1)(\ell+d-2)}{2\pi e^2 R^2})
+\frac{1}{2}\sum_{\ell=1}^{\infty} g_{\ell}^{(0)} \log(\frac{\ell(\ell+d-1)}{R^2})\,,\\
&F_{\rm min-sc}=\frac{1}{2}\log {\rm det}'(-\nabla^2) = \frac{1}{2}\sum_{\ell=1}^{\infty} g_{\ell}^{(0)} \log(\frac{\ell(\ell+d-1)}{R^2})\,,\\
&F_{\rm measure} = \log(2\pi  \sqrt{{\rm vol}(S^d)})\,.
\end{aligned}
\end{equation}
The grouping of terms in (\ref{F-feyn}) is essentially equivalent to doing the calculation in Feynman gauge, where one has an unconstrained
vector and a complex minimally coupled scalar ghost. To proceed, we use the identity
\begin{equation}
\log(y) = \int_0^{\infty} \frac{dt}{t} \left(e^{-t}-e^{-yt}\right)\,.
\label{heat}
\end{equation}
Then, using the dimensionally regularized identities (\ref{deg-sum}), one can rewrite the vector contribution as
\begin{equation}
F_{\rm vector} =-\frac{1}{2}\int_0^{\infty} \frac{dt}{t}
\left[\sum_{\ell=1}^{\infty} g_{\ell}^{(1)} (e^{-(\ell+1)t}+e^{-(\ell+d-2)t})+g_{\ell}^{(0)} (e^{-\ell t}+e^{-(\ell+d-1)t})\right]
-\frac{1}{2}\log(2\pi e^2)\,.
\end{equation}
Note that the radius dependence in $F_{\rm vector}$, and the terms proportional to $e^{-t}$, have dropped out due to (\ref{deg-sum}). The sum over $\ell$ can
now be evaluated analytically, leading to elementary functions of $e^{-t}$. To perform the $t$-integral, it is convenient to use the identity
\begin{equation}
\frac{1}{t} = \frac{1}{1-e^{-t}} \int_0^1 du e^{-u t}\,.
\label{t-To-u}
\end{equation}
This allows for an analytical evaluation of the $t$ integral, and after some algebra and using gamma function identities such as
$\Gamma(x)\Gamma(1-x) = \pi \csc(\pi x)$, we arrive at the result
\begin{eqnarray}
&&F_{\rm vector} = \int_0^1 du \bigg[
(d^2+1-3d(1+u)+2u(u+2))\sin(\frac{\pi}{2}(d-2u))\frac{\Gamma\left(d-2-u\right)\Gamma\left(1+u\right)}{2\sin(\frac{\pi d}{2})\Gamma\left(d\right)}
\cr
&&~~~~~~~~~~~~~-\frac{d-2}{(d-2)^2-u^2}\bigg]-\frac{1}{2}\log(2\pi e^2)\,.
\label{F-vec}
\end{eqnarray}
To evaluate the ghost contribution $F_{\rm min-sc}$ by similar methods, we can introduce a small regulator to deal with the zero mode, so that
we can extend the sum over all modes and make use of (\ref{deg-sum})
\begin{equation}
F_{\rm min-sc} = \lim_{\delta\rightarrow 0} \left[-\frac{1}{2}\int_0^{\infty} \frac{dt}{t}
\sum_{\ell=0}^{\infty} g_{\ell}^{(0)} (e^{-(\ell+\delta)t}+e^{-(\ell+d-1)t})-\frac{1}{2}\log(\frac{\delta (d-1)}{R^2})\right]\,.
\end{equation}
Performing first the sum over $\ell$, using (\ref{t-To-u}) and evaluating the $t$-integral, we obtain, after sending $\delta \rightarrow 0$ at
the end\footnote{We use $\log(\delta) = -\int_0^1 du \frac{1}{u+\delta}+\log(1+\delta)$.}
\begin{equation}
F_{\rm min-sc} = -\int_0^1 du \left[(d-2u)\sin(\frac{\pi}{2}(d-2u))
\frac{\Gamma\left(d-u\right)\Gamma\left(u\right)}{2\sin(\frac{\pi d}{2})\Gamma\left(d+1\right)}
-\frac{1}{2u}\right]-\frac{1}{2}\log(\frac{(d-1)}{R^2})\,.
\label{F-gh}
\end{equation}

We can now put everything together in (\ref{F-feyn}) and obtain the radius independent part of the Maxwell free energy (\ref{FMax}).
We find
\begin{equation}
F_{\rm Max.}^{(0)}(d) = \frac{1}{2}\log\left(2\pi (d-1)^2\Omega_d\right)-\frac{1}{\sin(\frac{\pi d}{2})}\int_0^1 du f_d(u)\,,
\label{FMax0}
\end{equation}
where the form of $f_d(u)$ can be read off
from the above results, and it is equal to
\begin{eqnarray}
&&\!\!\!\!\!\!\!\!\!\!\!\!\!\!\!\!\!\!
f_d(u) = -(d^2+1-3d(1+u)+2u(u+2))\sin(\frac{\pi}{2}(d-2u))\frac{\Gamma\left(d-2-u\right)\Gamma\left(1+u\right)}{2\Gamma\left(d\right)}
+\frac{\sin(\frac{\pi d}{2})(d-2)}{(d-2)^2-u^2}\cr
&&~~~~~~~~~~
-(d-2u)\sin(\frac{\pi}{2}(d-2u))
\frac{\Gamma\left(d-u\right)\Gamma\left(u\right)}{\Gamma\left(d+1\right)}
+\frac{\sin(\frac{\pi d}{2})}{u} \,.
\end{eqnarray}
Here the first line comes from the vector contribution (\ref{F-vec}), and the second line from the ghost contribution (\ref{F-gh}). Note that the
the UV divergences of the free energy are fully accounted for by the overall sine factor in front of the integral in (\ref{FMax0}).

Equivalently, in terms of $\tilde F$ we have
\begin{equation}
\tilde F_{\rm Maxwell} = \frac{1}{2}\sin(\frac{\pi d}{2})\log (e^2 R^{4-d})
-\frac{1}{2}\sin(\frac{\pi d}{2})\log\left(2\pi (d-1)^2\Omega_d\right)
+\int_0^1 du f_d(u)\,,
\label{tF}
\end{equation}
which is a finite smooth function of continuous $d$.

As a test of this result, we can check that in $d=4$ $\tilde F$ reproduces the known value of the conformal anomaly $a$-coefficient for the Maxwell theory.
From (\ref{tF}), we obtain
\begin{equation}
\tilde F_{\rm Maxwell}^{d=4} = \frac{\pi}{12} \int_0^1 du (1-u) (u^3-u^2-11 u+12) = \frac{\pi}{2}\cdot \frac{31}{45}\,
\end{equation}
corresponding to the correct $a$ anomaly coefficient, $a=\frac{31}{45}$ (we use units where $a=\frac{1}{90}$ for a 4d conformal scalar field).

In other even values of $d$, the Maxwell theory is not conformal and $\tilde F$ cannot be interpreted as an anomaly coefficient. Nevertheless,
$\tilde F$ still yields the coefficient of the $1/\epsilon$ pole in dimensional regularization, which fixes
the coefficient of the curvature counterterm in the renormalized free energy. From (\ref{tF}), we find for instance
\begin{equation}
\tilde F_{\rm Maxwell}^{d=6} = -\frac{\pi}{2}\cdot \frac{1271}{1890}\,,\quad
\tilde F_{\rm Maxwell}^{d=8} = \frac{\pi}{2}\cdot \frac{4021}{6300}\,,\quad
\tilde F_{\rm Maxwell}^{d=10} = -\frac{\pi}{2}\cdot \frac{456569}{748440}\,,\ldots\,.
\label{a-Maxwell}
\end{equation}
The $d=6$ result agrees with the value obtained in Appendix of \cite{Beccaria:2015uta}. For other even $d$ values, we have checked
that our results are in agreement with the coefficient of the logarithmic divergence for a massless spin 1 field
obtained by zeta function methods on Euclidean $AdS_{2n}$ \cite{Camporesi:1994ga}.

As a further check, in $d=3$ we obtain the result
\begin{eqnarray}
F_{\rm Maxwell}^{d=3} &=& -\frac{1}{2} \log \left(\frac{e^2 R}{16 \pi ^3}\right)
-\int_0^1 du \left[\frac{1}{1-u^2}+\frac{1}{u}-\frac{\pi}{12}(2 u^3+3 u^2-23 u+12) \cot (\pi  u) \right]\cr
&=& -\frac{1}{2}\log(e^2 R) +\frac{\zeta(3)}{4\pi^2}
\end{eqnarray}
in agreement with \cite{Klebanov:2011td}. In $d=3$, the Maxwell theory is Hodge dual to a compact minimally coupled scalar field. Note that
from (\ref{F-gh}) we can read off the $F$-value for a (non-compact) minimal scalar in $d=3$, with zero mode removed, to be
\begin{eqnarray}
F_{\rm min-sc}^{d=3} = \frac{1}{2}\log(\pi)+\frac{\zeta(3)}{4\pi^2}+\log(R)\,.
\end{eqnarray}
This result agrees with the one obtained in \cite{Klebanov:2011td,Agon:2013iva}, and after carefully relating the radius of the compact scalar to the electric charge $e$,
one can verify equality of the partition functions under Hodge duality.

In $d=5$, we find for $\tilde F=-F$:
\begin{eqnarray}
\!\!\!
\tilde{F}_{\rm Maxwell}^{d=5} &=& \frac{1}{2} \log \left(\frac{e^2}{32\pi ^4 R}\right)
+\int_0^1 du \left[\frac{1}{u}-\frac{3}{u^2-9}-\frac{\pi}{240}  \left(6 u^5-35 u^4+275 u^2-486 u+240\right) \cot (\pi  u)\right]\cr
&=&\frac{1}{2} \log \left(\frac{e^2}{4 \pi ^2 R}\right)+\frac{5 \zeta (3)}{16 \pi ^2}+\frac{3 \zeta (5)}{16 \pi ^4}\,.
\end{eqnarray}
It would be interesting to reproduce this result from a massless 2-form $B_2$ on $S^5$, which is related by Hodge duality to the Maxwell theory.

\section{Conformal QED at large $N$}
\label{largeN}

The action for Maxwell theory coupled to $N_f$ massless charged fermions in flat space is (in Euclidean signature)
\begin{equation}
S = \int d^d x \left(\frac{1}{4e^2}F^{\mu\nu}F_{\mu\nu} -\sum_{i=1}^{N_f} \bar\psi_i \gamma^{\mu}(\partial_{\mu}+i A_{\mu})\psi^i \right)\,.
\end{equation}
Here the fermions $\psi^i$ are assumed to be {\it four-component} complex spinors. These correspond to $N_f$ usual Dirac fermions in $d=4$,
while in $d=3$ they can be viewed as $2N_f$ 3d Dirac fermions. In particular, in $d=3$ the model has the enhanced
flavor symmetry $SU(2N_f)$. We define the dimensional continuation of the theory by keeping
the number of fermion components fixed. In other words, we take $\gamma^{\mu}$ to be $4\times 4$ matrices satisfying $\{\gamma^{\mu},\gamma^{\nu}\} = 2\delta^{\mu\nu}\, {\bf 1}$, with $\tr {\bf 1} = 4$. All vector indices are formally continued to $d$ dimensions, i.e. $g^{\mu\nu}g_{\mu\nu}=d$, $\gamma^{\mu}\gamma_{\mu} = d\cdot{\bf 1}$, etc.

One may develop the $1/N$ expansion of the theory by integrating out the fermions. This produces an effective action for the gauge field of the form
\begin{equation}
S_{\rm eff} = \int d^d x \frac{1}{4e^2}F^{\mu\nu}F_{\mu\nu}-\int d^d x d^d y\left(\frac{1}{2} A^{\mu}(x)A^{\nu}(y)\langle J_{\mu}(x)J_{\nu}(y)\rangle_0+
\mathcal{O}(A^3)\right)\,,
\label{Seff}
\end{equation}
where
\begin{equation}
J_{\mu} = i \bar\psi_i \gamma_{\mu}\psi^i
\end{equation}
is the conserved $U(1)$ current. Using the fermion propagator
\begin{equation}
\langle \psi^i(x)\bar\psi_j(0)\rangle =
-\delta^i_j \frac{\Gamma\left(\frac{d}{2}\right)}{2\pi^{\frac{d}{2}}}\frac{\gamma^{\mu}x_{\mu}}{(x^2)^{\frac{d}{2}}}
= i\delta^i_j \int \frac{d^dp}{(2\pi)^d}  \frac{\gamma^{\mu}p_{\mu}}{p^2}\,e^{i p x}
\end{equation}
the current two-point function in the free fermion theory is found to be
\begin{equation}
\langle J_{\mu}(x)J_{\nu}(0)\rangle_0 = C_J \frac{g_{\mu\nu}-2\frac{x_{\mu}x_{\nu}}{x^2}}{x^{2(d-1)}}\,,\qquad
C_J 
= N_f \tr  {\bf 1} \left(\frac{\Gamma\left(\frac{d}{2}\right)}{2\pi^{\frac{d}{2}}}\right)^2\,.
\end{equation}
In momentum space, one finds\footnote{More generally, for a spin 1 primary operator of dimension $\Delta$, one has
$\langle J_{\mu}(x)J_{\nu}(0)\rangle = C_J \frac{g_{\mu\nu}-2\frac{x_{\mu}x_{\nu}}{x^2}}{x^{2\Delta}}$ and
$\langle J_{\mu}(p)J_{\nu}(-p)\rangle=C_J \frac{2^{d-2\Delta} \pi ^{d/2} (\Delta-1)\Gamma \left(\frac{d}{2}-\Delta\right)}{\Gamma (\Delta+1)}
\left(g_{\mu\nu}-\frac{2\Delta-d}{\Delta-1}\frac{p_{\mu}p_{\nu}}{p^2}\right)(p^2)^{\Delta-\frac{d}{2}}$.}
\begin{equation}
\langle J_{\mu}(p)J_{\nu}(-p)\rangle_0 = \int d^dx e^{-i p x} \langle J_{\mu}(x)J_{\nu}(0)\rangle_0=
-C_J \frac{2^{3-d} \pi ^{d/2} \Gamma \left(2-\frac{d}{2}\right)}{\Gamma (d)}\left(g_{\mu\nu}-\frac{p_{\mu}p_{\nu}}{p^2}\right)(p^2)^{\frac{d}{2}-1}\,.
\label{A2-induced}
\end{equation}
Thus, when $d<4$, one sees that the non-local kinetic term in (\ref{Seff}) is dominant in the low momentum (IR) limit compared to the two-derivative
Maxwell term. Hence, the latter can be dropped at low energies, and one may develop the $1/N$ expansion of the critical theory by
using the induced quadratic term
\begin{equation}
S_{\rm crit} = -\frac{1}{2}\int \frac{d^dp}{(2\pi)^d} A^{\mu}(p)A^{\nu}(-p)\langle J_{\mu}(p)J_{\nu}(-p)\rangle_0+\mathcal{O}(1/N_f)\,.
\end{equation}
Note that this effective action is gauge invariant as it should, due to conservation of the current.

To compute the sphere free energy, we need to conformally map to $S^d$ and choose an appropriate gauge fixing.
As in the previous section, we may gauge fix by splitting $A_{\mu}=B_{\mu}+\partial_{\mu}\phi$, where $\nabla_{\mu}B^{\mu}=0$. Then, following the
same steps as in (\ref{Zdef}), (\ref{ZMax}), the sphere free energy is given by
\begin{equation}
F = N_f F_{\rm free-ferm}(d)+\frac{1}{2}\log{\rm det}_{T}\left(\frac{K_{\mu\nu}}{2\pi}\right)
-\frac{1}{2}\log{\rm \det}'(-\nabla^2)+\log\left(2\pi \sqrt{{\rm vol}(S^d)}\right)
+\mathcal{O}(\frac{1}{N_f})\,,
\label{F-largeN}
\end{equation}
where $K_{\mu\nu}=-\langle J_{\mu}J_{\nu}\rangle$ is the non-local induced kinetic term, and $F_{\rm free-ferm}$
is the contribution of a free
four-component Dirac fermion \cite{Giombi:2014xxa}
\begin{equation}
F_{\rm free-ferm}(d) = -\frac{4}{\sin(\frac{\pi d}{2})\Gamma\left(1+d\right)}\int_0^{1} du\, \cos\left(\frac{\pi u}{2}\right)\Gamma\left(\frac{1+d+u}{2}\right)\Gamma\left(\frac{1+d-u}{2}\right)\,.
\label{Ffer}
\end{equation}
The ghost contribution was already computed in the previous section, and is given in (\ref{F-gh}).
To evaluate the contribution of the transeverse vector, we first conformally map the current two-point function to the sphere of radius $R$, on which we choose
the conformally flat metric
\begin{equation}
ds^2= \frac{4R^2 dx^{\mu}dx^{\mu}}{(1+x^2)^2}\,.
\end{equation}
Introducing the vielbein $e^m_{\mu}(x) = \frac{2R}{(1+x^2)}\delta^{m}_{\mu}$, the two-point function for a spin 1 primary operator of dimension $\Delta$ can be written as
\begin{equation}
\langle J_{\mu}(x) J_{\nu}(y)\rangle = C_J e_{\mu}^m(x) e_{\nu}^n(y) \frac{\left(\delta^{mn}-2\frac{(x-y)^m (x-y)^n}{|x-y|^2}\right)}{s(x,y)^{2\Delta}}\,,\qquad
s(x,y) =\frac{2R|x-y|}{(1+x^2)^{1/2}(1+y^2)^{1/2}}\,,
\end{equation}
where in our case $\Delta=d-1$, corresponding to a conserved current. The spin 1 determinant in (\ref{F-largeN}) may be computed by expanding in
a basis of vector spherical harmonics \cite{Gubser:2002vv, Klebanov:2011td, Giombi:2013yva}. Splitting the vector $A_{\mu}$ in transverse and longitudinal parts,
the spin 1 and spin 0 eigenvalues of $K_{\mu\nu}=-\langle J_{\mu} J_{\nu}\rangle$ turn out to be, in the case of general conformal dimension $\Delta$ (see Appendix \ref{AppK}):
\begin{equation}
\begin{aligned}
&\lambda_{\ell}^{(1)} = -C_J \frac{2^{d-2\Delta} \pi ^{d/2} (\Delta-1)\Gamma \left(\frac{d}{2}-\Delta\right)}{\Gamma (\Delta+1)}
\frac{\Gamma\left(\ell+\Delta\right)}{\Gamma\left(\ell+d-\Delta\right)}\frac{1}{R^{2\Delta-d}}\,,\\
&\lambda_{\ell}^{(0)} = \frac{d-1-\Delta}{\Delta-1}\lambda_{\ell}^{(1)}\,, \label{lambs}
\end{aligned}
\end{equation}
with degeneracies given in (\ref{deg-1}) and (\ref{deg-0}). For $\Delta=d-1$ the longitudinal eigenvalues vanish as expected, due to gauge invariance. The
spin 1 contribution in (\ref{F-largeN}) is then
\begin{equation}
\frac{1}{2}\log{\rm det}_{T}\left(\frac{K_{\mu\nu}}{2\pi}\right) =
\frac{1}{2} \log \left(N_f\frac{ \Gamma \left(2-\frac{d}{2}\right)\Gamma \left(\frac{d}{2}\right)^2}
{2^{d-2} \pi^{\frac{d}{2}+1}\Gamma \left(d\right)R^{d-2}}\right)+\frac{1}{2}\sum_{\ell=1}^{\infty}
g_{\ell}^{(1)} \log\left(\frac{\Gamma\left(\ell+d-1\right)}{\Gamma\left(\ell+1\right)}\right)\,,
\end{equation}
where we have used the dimensionally regularized identity (\ref{deg-sum}) to extract the constant prefactor in the eigenvalues, and used
$C_J=4N_f \left(\frac{\Gamma\left(\frac{d}{2}\right)}{2\pi^{\frac{d}{2}}}\right)^2$. From this expression, we immediately see that the free
energy contains a term $\frac{1}{2}\log(N_f)$, independently of dimension. This can be traced back to the trivial constant gauge transformations
on the sphere, or equivalently to ghost zero modes \cite{Giombi:2013yva}. Note also that the radius dependence cancels out against the ghost and measure
contributions in (\ref{F-largeN}), as expected by conformal invariance. The remaining non-trivial sum may be evaluated directly
for instance by using the integral representation
\begin{equation}
\log\Gamma(z) = \int_0^{\infty} dt\left(z-1-\frac{1-e^{-(z-1)t}}{1-e^{-t}}\right)\frac{e^{-t}}{t}
\label{logGamma}
\end{equation}
and following similar steps as described in the previous section.
A compact form of the final answer for the sum is suggested by the results of \cite{Giombi:2013yva}, where a formula for the change in $F$ due to a deformation by the
square of a spin $s$ operator of dimension $\Delta$ was computed using higher spin fields in $AdS_{d+1}$ with non-standard boundary
conditions. For spin 1, that result implies:
\begin{eqnarray}
&&\frac{1}{2}\sum_{\ell=1}^{\infty} g_{\ell}^{(1)}\log\left(\frac{\Gamma\left(\ell+\Delta\right)}{\Gamma\left(\ell+d-\Delta\right)}\right)
+\frac{1}{2}\sum_{\ell=1}^{\infty} g_{\ell}^{(0)}
\log\left(\frac{d-1-\Delta}{\Delta-1}\frac{\Gamma\left(\ell+\Delta\right)}{\Gamma\left(\ell+d-\Delta\right)}\right)\cr
&&~~~~=-\frac{1}{\sin\left(\frac{\pi d}{2}\right)\Gamma\left(d\right)}
\int_0^{\Delta-\frac{d}{2}} du\, u(d^2-4u^2)\sin(\pi u)\Gamma\left(\frac{d}{2}-1+u\right)\Gamma\left(\frac{d}{2}-1-u\right)\,.
\end{eqnarray}
Taking carefully the limit of $\Delta=d-1$, and using \cite{Diaz:2007an,Giombi:2014xxa} (note that the sum starts from $\ell=0$ here):
\begin{equation}
\frac{1}{2}\sum_{\ell=0}^{\infty} g_{\ell}^{(0)}
\log\left(\frac{\Gamma\left(\ell+\Delta\right)}{\Gamma\left(\ell+d-\Delta\right)}\right)
=-\frac{1}{\sin\left(\frac{\pi d}{2}\right)\Gamma\left(d+1\right)}
\int_0^{\Delta-\frac{d}{2}} du\, u \sin(\pi u) \Gamma\left(\frac{d}{2}+u\right)\Gamma\left(\frac{d}{2}-u\right)
\end{equation}
we finally obtain the result
\begin{equation}
\begin{aligned}
&\frac{1}{2}\sum_{\ell=1}^{\infty}
g_{\ell}^{(1)} \log\left(\frac{\Gamma\left(\ell+d-1\right)}{\Gamma\left(\ell+1\right)}\right)=
\frac{1}{2}\log\left (\frac {\Gamma\left(d-1\right)}{2}\right ) \\
&-\int_0^1 du\bigg[(d-2)^2 (d-1)u\left(4+d^2-(d-2)^2 u^2\right)\frac{\sin \left(\frac{\pi  (d-2) u}{2} \right)
\Gamma  \Big(\frac{(d-2)(1-u)}{2} \Big) \Gamma \Big(\frac{(d-2)(1+u)}{2}\Big)}{16\sin \left(\frac{\pi  d}{2}\right)\Gamma (d+1)}+
\frac{1}{2(1-u)}\bigg]\,.
\end{aligned}
\label{Spin1Sum}
\end{equation}
We have explicitly verified that this agrees with a direct evaluation of the sum using (\ref{logGamma}).

Putting everything together, the final result for the sphere free energy $F$, or equivalently for $\tilde F=-\sin(\frac{\pi d}{2})F$,
takes the form\footnote{Note that, due to the factor $\log \left (-N_f/\sin(\frac{\pi d}{2})\right )$, the free energy is real for $2\le d \le 4$, it has an imaginary part for $4<d<6$, then it is real again for $6 \le d \le 8$, etc. This is essentially due to the fact that the Maxwell term yields a contribution $-\frac{1}{2}\log(e^2 R^{4-d})$ to $F$, and at the RG fixed point $e_*^2$ is positive for $2 < d < 4$, negative for $4<d<6$, etc.} 
\begin{equation}
\tilde F =N_f \tilde{F}_{\rm free-ferm}(d)-\frac{1}{2}\sin(\frac{\pi d}{2})\log \left (-\frac{N_f}{\sin(\frac{\pi d}{2})}\right )+A_0(d)+\mathcal{O}(\frac{1}{N_f})\,,
\label{tF-largeN}
\end{equation}
where
\begin{align}
A_0(d) =& -\sin(\frac{\pi d}{2})\bigg[
\frac{1}{2}\sum_{\ell=1}^{\infty} g_{\ell}^{(1)} \log\left(\frac{\Gamma\left(\ell+d-1\right)}{\Gamma\left(\ell+1\right)}\right)
-\frac{1}{2}\sum_{\ell=1}^{\infty} g_{\ell}^{(0)}\log\left(\ell(\ell+d-1)\right)\notag\\
&+
\frac{1}{2}\log\bigg(
\frac{2^{5-2 d}\pi ^3 (d-2)}{
\Gamma \left(\frac{d+1}{2}\right)^2}
\bigg)
\bigg]
\label{A0}
\end{align}
and the sums can be given the integral representations in (\ref{F-gh}) and (\ref{Spin1Sum}).
The resulting $A_0(d)$ is a smooth, finite function of $d$ which is independent of $R$ and $N_f$. In $d=3$ it evaluates to
\begin{equation}
A_0(d=3) =\frac{1}{2}\log\left(\frac{\pi}{4}\right)
\end{equation}
and so we find agreement with (\ref{confQED}).
For comparison to the perturbative calculation in the $\epsilon$ expansion given in the
next section, it is also useful to expand (\ref{A0}) in $d=4-\epsilon$. We find
\begin{equation}
A_0(d=4-\epsilon)=\frac{31\pi}{90}
-0.905 \epsilon - 0.64931 \epsilon^2 + 0.374025 \epsilon^3+\mathcal{O}(\epsilon^4)\,.
\label{largeN-4mep}
\end{equation}
The leading term correctly reproduces the $a$-anomaly coefficient of the $d=4$ Maxwell field, as expected. In the next section we
will reproduce the 
terms to order $\epsilon^3$ from a perturbative
calculation on $S^{4-\epsilon}$.

\begin{figure}[h!]
                \centering
                \includegraphics[width=12cm]{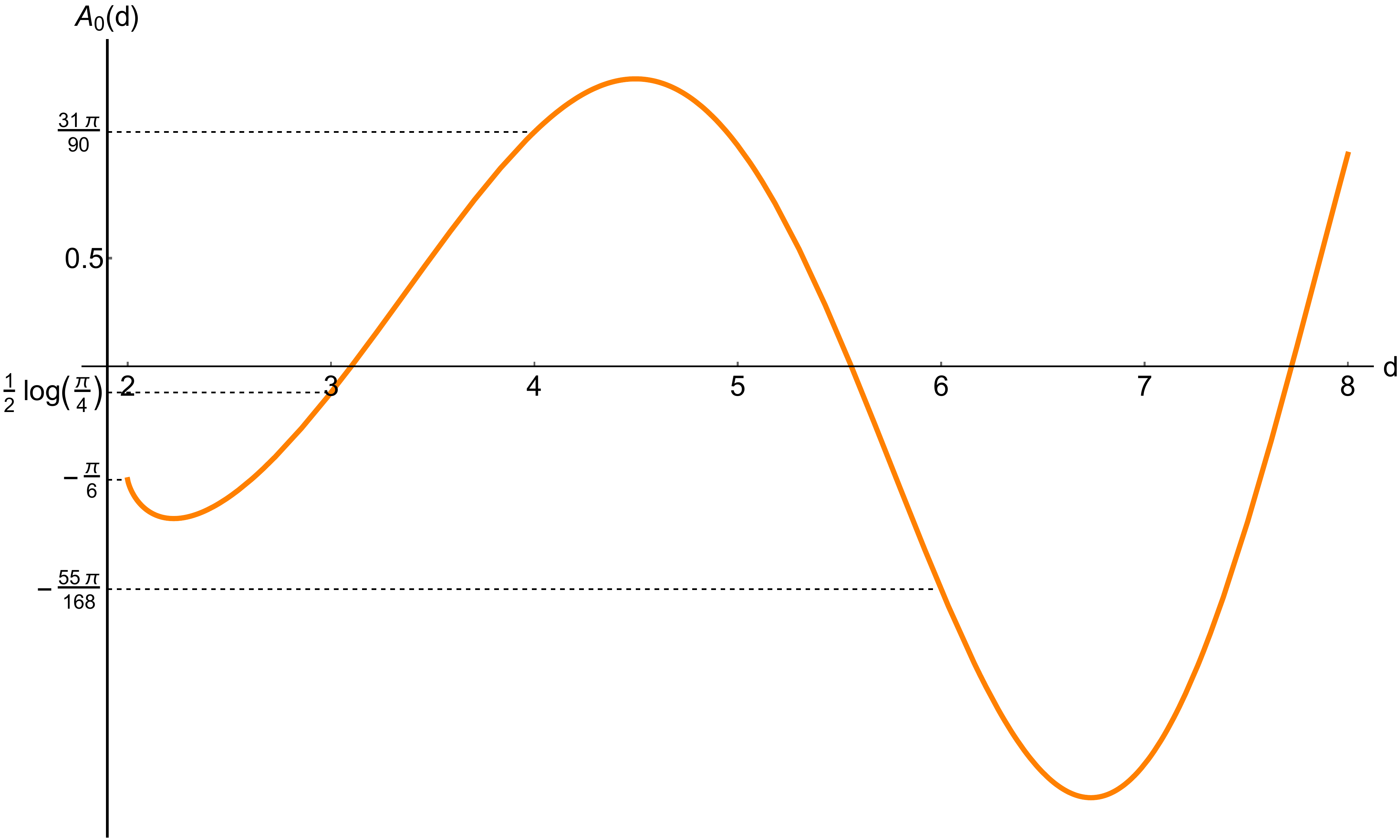}
                \caption{Plot of the smooth function $A_0(d)$
from eq.~(\ref{tF-largeN}).
It has values  $\tilde F=-55\pi/168$ ($a=-55/84$) in $d=6$,
$\tilde F=31\pi/90$ ($a=31/45$) in $d=4$, and $\tilde F = -\pi/6$ ($c=-1$) in $d=2$.}
                \label{A0d}
\end{figure}

Let us also note that in $d=2$ we find
\begin{equation}
A_0(d=2) = -\frac{\pi}{6}
\end{equation}
corresponding to a shift of the central charge by $-1$. This is as expected, since in $d=2$ we get the Schwinger model
coupled to $N_f$ massless fermions; via the non-abelian bosonization \cite{Witten:1983ar} one finds that at low energies it is a CFT with central charge $c=2N_f-1$ \cite{Gepner:1984au,Affleck:1985wa}.
This result is exact (all the $1/N_f$ corrections
to $\tilde F$ should vanish as $d\rightarrow 2$), and we will make use of it in Section \ref{PadeSec} to impose a boundary condition on the Pad\'e extrapolations
of our $\epsilon$ expansion results. A plot of the function $A_0(d)$
is given in Fig.~\ref{A0d}.

\subsection{Comments on $d>4$}
\label{comments}

In $d>4$, one still formally finds a conformal electrodynamics in the large momentum (UV) limit, see eq.~(\ref{A2-induced}),
but the corresponding CFT's are non-unitary. For instance, in $d=6$ the induced kinetic term (\ref{A2-induced}) corresponds
to the conformal spin 1 gauge field with Lagrangian $L \sim F_{\mu\nu}\partial^2F^{\mu\nu}$
\cite{Fradkin:1985am, Giombi:2013yva, Tseytlin:2013fca}. The $a$-anomaly coefficient for this conformal field can be extracted
from our general result (\ref{tF-largeN}) setting $d=6$, which yields
\begin{eqnarray}
(\tilde F - N_f {\tilde F}_{\rm free-ferm})|_{d=6} &=&
\frac{\pi}{240}\int_0^1 du \left(213 u^6+6 u^5-630 u^4+160 u^3-183 u^2+314 u-120\right)\cr
&=&-\frac{\pi}{2}\cdot \frac{55}{84}\,.
\end{eqnarray}
corresponding to $a=-\frac{55}{84}$ (in units where $a=\frac{1}{756}$ for a 6d conformal scalar).
This agrees with the result for the $a$-anomaly of the 6d conformal spin 1 field,
which can be obtained from one-loop determinants in $AdS_7$ with non-standard boundary conditions \cite{ Giombi:2013yva, Giombi:2014iua}, or by a direct computation on $S^6$ \cite{Tseytlin:2013fca}. Note
that this is {\it not} equal to the coefficient of the logarithmic divergence for a ordinary Maxwell field in $d=6$, eq.~(\ref{a-Maxwell}).
As recently observed in \cite{Beccaria:2015uta}, this conformal spin 1 field
is part of a non-unitary ${\cal N}=(1,0)$ conformal multiplet including a Weyl fermion with 3-derivative kinetic term
and 3 conformal scalar fields, whose total $a$ anomaly coefficient is $a=-\frac{251}{360}$,
which turns out to be the value assigned by definition in \cite{Cordova:2015fha} to a ${\cal N}=(1,0)$ vector multiplet in $d=6$.

For finite $N_f$, one approach to the conformal QED in $d>4$ is to use the $d=4+\epsilon$ expansion. From the one-loop beta function (\ref{beta-QED-oneloop}),
one sees that there are UV fixed points at imaginary values of the coupling. The large $N_f$ limit considerations discussed above strongly suggest
that these UV fixed points have a UV completion in $d=6-\epsilon$ as the IR fixed points of the higher derivative renormalizable gauge theory
\begin{equation}
S = \int d^d x \left(\frac{1}{4 e_0^2} F_{\mu\nu}(-\nabla^2)F^{\mu\nu}-\bar\psi_i \gamma^{\mu}(\partial_{\mu}+i A_{\mu})\psi^i\right)\,,
\end{equation}
where $\psi^i$ are $N_f$ 6d Weyl fermions. To get an anomaly free theory, we may add $N_f$ Weyl fermions 
of the opposite chirality, so that the model includes $N_f$ 6d Dirac fermions.  
The one-loop beta function for this theory can be computed by evaluating the
correction to the gauge field propagator due to the fermion loop, which is given by (\ref{A2-induced}) for general $d$. Expanding in $d=6-\epsilon$,
one finds a pole that fixes the charge renormalization, and for the theory with $N_f$ Dirac fermions, we obtain the beta function
\begin{equation}
\beta_e = -\frac{\epsilon}{2}e-\frac{N_f}{120\pi^3}e^3+\mathcal{O}(e^5)\,.
\end{equation}
Unlike the case of QED$_4$, this theory is asymptotically free in $d=6$.
It would be interesting to compute the beta function
for the non-abelian version of this model. By analogy with $d=4$, we expect that in this case the pure glue should give a {\it positive} contribution
to the beta function, while matter gives negative contributions. The pure glue theory may then have IR fixed points for positive $g^2$ that could provide a UV completion of the Yang-Mills theory in $d=4+\epsilon$. Also, in the theory with gauge group $SU(N_c)$ and $N_f$ massless fermions, one may contemplate
the existence of a conformal window directly in $d=6$.

\section{Sphere free energy in the $\epsilon$ expansion}
\label{F-eps}

The action for massless QED in $d=4-\epsilon$ on a general curved Euclidean manifold is
\begin{align}
S= \int d^{d}x \sqrt{g_{x}} \Big(\frac{1}{4e_0^2}F_{\mu\nu}F^{\mu\nu}
+\frac{1}{2}(\nabla_{\mu}A^{\mu})^{2}
-\sum_{i=1}^{N_f}\bar{\psi}_{i}\gamma^{\mu} (\nabla_{\mu} +i A_{\mu})\psi^{i}
+a_{0}W^2+b_{0}E+c_{0}\mathcal{R}^{2}/(d-1)^2\Big)\,,
\label{QED-curved}
\end{align}
where $\psi^i$ are $N_f$ four-component Dirac fermions, and we have added a Feynman gauge fixing term, which we find most convenient for the perturbative
calculation below. Here $\nabla_{\mu}$ is the curved space covariant derivative (when it acts on fermions, it includes the spin connection term as usual).
Finally, $\mathcal{R}$ denotes the Ricci scalar, $W^2$ is the square of the Weyl tensor and $E$ is the Euler density:
\begin{align}
&W^2= \mathcal{R}_{\mu\nu\lambda\rho}  \mathcal{R}^{\mu\nu\lambda \rho} - \frac{4}{d-2}  \mathcal{R}_{\mu\nu} \mathcal{R}^{\mu\nu} + \frac{2}{(d-2)(d-1)} \mathcal{R}^{2},
\notag\\
& E=  \mathcal{R}_{\mu\nu\lambda\rho}  \mathcal{R}^{\mu\nu\lambda \rho} -4 \mathcal{R}_{\mu\nu} \mathcal{R}^{\mu\nu} + \mathcal{R}^{2}.
\label{curv-conv}
\end{align}
The action includes all terms that are marginal in $d=4$, and $e_0, a_0, b_0, c_0$ are the corresponding bare coupling parameters. Renormalizability
of the theory on an arbitrary manifold implies that the divergencies of the free energy can be removed by a suitable renormalization of the
bare parameters which is independent of the background metric. The renormalization of the electric charge is fixed by the flat space theory and reads,
in minimal subtraction scheme \cite{Moshe:2003xn}:
\begin{equation}
e_{0} = \mu^{\frac{\epsilon}{2}}\bigg(e+\frac{4  N_{f}}{3  \epsilon }\frac{e^{3}}{(4 \pi )^2}
+\Big(\frac{8  N_{f}^{2}}{3 \epsilon^{2}}+\frac{2 N_{f}}{\epsilon}\Big)\frac{e^5}{(4 \pi )^4}+
\left(\frac{160 N_f^3}{27 \epsilon^3}+\frac{88 N_f^2}{9 \epsilon^2}-\frac{2N_f (22N_f+9)}{27 \epsilon}\right)\frac{e^7}{(4\pi)^6}+\ldots \bigg)\,,
\label{e0Toe}
\end{equation}
where $e$ is the renormalized coupling, and the corresponding beta function is\footnote{The terms of order $e^9$ that we have omitted from (\ref{e0Toe}) can
be reconstructed from (\ref{beta-e}) if desired.}
\begin{equation}
\beta=-\frac{\epsilon }{2}e+\frac{4N_{f}}{3}\frac{e^3}{ (4 \pi )^2}+\frac{4 N_{f} e^5 }{(4 \pi )^4}-\frac{2 N_{f} (22 N_{f}+9)}{9 }\frac{e^7}{(4 \pi )^6}-\frac{2 N_{f} (4N_{f} (154 N_{f}+2808 \zeta (3)-855)+5589)e^9}{243 (4 \pi )^8}\,.
\label{beta-e}
\end{equation}
Then, one finds an IR stable perturbative fixed point at
\begin{equation}
e_{*}=\pi  \sqrt{\frac{6 \epsilon }{N_{f}}} \Big(1-\frac{9 }{16 N_{f}}\epsilon +\frac{3 (44 N_{f}+207) }{512 N_{f}^2}\epsilon ^2+\frac{ (2464 N_{f}^2+44928 N_{f} \zeta (3)-45756 N_{f}-62937)}{24576 N_{f}^3}\epsilon ^3+\mathcal{O}(\epsilon^4)\Big)\,.
\label{estar}
\end{equation}

The first few terms in the renormalization of the curvature couplings have been obtained in \cite{Hathrell:1981gz} for $N_f=1$,
and in \cite{Jack:1990eb} for the general case. In our conventions, they read\footnote{The term of order $e^6/\epsilon$ in $b_0$
is scheme dependent in the sense that it depends on the definition of the Euler density $E$ in $d=4-\epsilon$. Our conventions for $E$ in (\ref{curv-conv})
differ from \cite{Jack:1990eb} by an overall $d$-dependent factor. One can verify that the free energy at the fixed point is not affected
by this convention dependence.}
\begin{equation}
\begin{aligned}
&a_0 = \mu^{-\epsilon}\left(a+\frac{N_f+2}{20\epsilon(4\pi)^2}+\frac{7N_f}{72\epsilon}\frac{e^2}{(4\pi)^4}+\ldots\right)\,,\\
&b_0 = \mu^{-\epsilon}\left(b-\frac{11N_f+62}{360\epsilon(4\pi)^2}+\frac{ N_{f}}{6  \epsilon } \frac{e^4}{(4 \pi )^6}
+\Big(\frac{2  N_{f}^2}{9 \epsilon ^2}-\frac{(16  N_{f}+9) N_{f}}{108 \epsilon }\Big) \frac{e^6}{(4 \pi )^8}+\ldots\right)\,,\\
&c_0 = \mu^{-\epsilon}\left(c-\frac{ N_{f}^2}{9 \epsilon } \frac{ e^6}{(4 \pi )^8 }+\ldots\right)\,.
\label{abc-ren}
\end{aligned}
\end{equation}
and the corresponding beta functions for the renormalized parameters $a, b, c$ are
\begin{equation}
\begin{aligned}
&\beta_a = \epsilon a +\frac{N_f+2}{20(4\pi)^2}+\frac{7N_f}{36}\frac{e^2}{(4\pi)^4}+\ldots\,,\\
&\beta_b =\epsilon b-\frac{11N_f+62}{360(4\pi)^2}+\frac{ N_{f}}{2 } \frac{e^4}{(4 \pi )^6}
-\frac{(16  N_{f}+9) N_{f}}{27} \frac{e^6}{(4 \pi )^8}+\ldots\,, \\
&\beta_c=\epsilon c-\frac{4N_{f}^2}{9} \frac{ e^6}{(4 \pi )^8 }+\ldots\,.
\end{aligned}
\end{equation}

We are interested in computing the free energy of the theory on a round sphere $S^d$ of radius $R$, for which
one has $W^2=0, E=d(d-1)(d-2)(d-3)/R^4, {\mathcal R}=d(d-1)/R^2$. In particular, the renormalization of the Weyl square coupling $a_0$
will not play any role in this calculation. After renormalization, the sphere free energy $F(e,b,c,\mu R)$ is a finite function for any value
of the renormalized couplings $e, b, c$. By standard arguments, it satisfies the Callan-Symanzik equation
\begin{equation}
\left(\mu \frac{\partial}{\partial\mu}+\beta_{e}\frac{\partial}{\partial e}+
\beta_b \frac{\partial}{\partial b}+
\beta_c \frac{\partial}{\partial c}\right)F(e,b,c,\mu R)=0\,.
\label{CS}
\end{equation}
As explained in \cite{Fei:2015oha}, it follows that to obtain the radius independent free energy at the IR fixed point we should set not only $\beta_e=0$, but also
the curvature beta functions $\beta_b=\beta_c=0$. The corresponding fixed point values in $d=4-\epsilon$ are given by $e=e_*$ in eq.~(\ref{estar}), and
\begin{equation}
\begin{aligned}
&b_* = \frac{1}{\epsilon}\left(\frac{11N_f+62}{360(4\pi)^2}-\frac{ N_{f}}{2 } \frac{e_*^4}{(4 \pi )^6}
+\frac{(16  N_{f}+9) N_{f}}{27} \frac{e_*^6}{(4 \pi )^8}\right)+\mathcal{O}(\eps^4)\,,\\
&c_* = \frac{1}{\epsilon}\,\frac{4N_{f}^2}{9} \frac{ e_*^6}{(4 \pi )^8 }+\mathcal{O}(\eps^4)\,,
\label{bc-star}
\end{aligned}
\end{equation}
and the free energy at the fixed point is the radius independent quantity
\begin{equation}
F_{\rm conf}(\epsilon) = F(e_*, b_*, c_*, \mu R)\,.
\end{equation}
Note that at the free field level, the effect of (\ref{bc-star}) is simply to remove the coupling independent part of the curvature terms. The
fixed point free energy has then a pole due to the free field determinants, but $\tilde F=-\sin(\pi d/2)F$ is finite in
the $d\rightarrow 4$ limit and reproduces the $a$ anomaly of the 4d theory \cite{Giombi:2014xxa, Fei:2015oha}.

To perform the explicit calculations, we find it convenient to follow \cite{Adler:1972qq, Adler:1973ty, Drummond:1977uy,  Shore:1977df} and describe the sphere by flat embedding
coordinates $\eta^a$, $a=1,2,\ldots,d+1$ satisfying $\sum_{a,b}\delta_{ab}\eta^a \eta^b = R^2$. In this approach, one also
introduces Dirac matrices $\alpha_a$ of dimension $2^{d/2}$ satisfying the Clifford algebra in $d+1$ dimensions $\{\alpha_{a},\alpha_{b}\}=2\delta_{ab}$, $a,b=1,...,d+1$. In
this embedding formalism, the vertex in (\ref{QED-curved}) is given by
\begin{align}
\Gamma_{a}(\eta) =ie_{0}Q_{ab}(\eta)\alpha_{b},\qquad  Q_{ab}(\eta) = \delta_{ab} - \frac{\eta_{a}\eta_{b}}{R^{2}}\,, \label{vertex}
\end{align}
where we have rescaled the gauge field to bring the coupling constant in the vertex. One advantage of using embedding coordinates is
that the propagators take a relatively simple form \cite{Drummond:1977uy}. The photon propagator in the Feynman gauge is
\begin{equation}
D_{ab}(\eta_{1},\eta_{2})  =\delta_{ab}D(\eta_{1},\eta_{2})= \delta_{ab} \frac{\Gamma(d-2)}{(4\pi)^{\frac{d}{2}}R^{d-2}\Gamma(\frac{d}{2})} \,_{2}F_{1}\left(1,d-2,\frac{d}{2},1-\frac{(\eta_{1}-\eta_{2})^{2}}{4R^{2}}\right)
\label{propA}
\end{equation}
and the fermion propagator is
\begin{equation}
S^{i}_{j}(\eta_{1},\eta_{2}) = -\delta^{i}_{j}\frac{\Gamma(\frac{d}{2})}{2\pi^{\frac{d}{2}}} \frac{\alpha \cdot (\eta_{1}-\eta_{2})}{|\eta_{1}-\eta_{2}|^{d}}.
\label{proppsi}
\end{equation}

Introducing the integrated $n$-point functions
\begin{align}
G_{n} = \int \prod_{k=1}^{n}d^{d}\eta_{k} \langle \bar{\psi}_{i_{1}}\Gamma_{a_{1}}A^{a_{1}}\psi^{i_{1}}(\eta_{1})\dots \bar{\psi}_{i_{n}}\Gamma_{a_{n}}A^{a_{n}}\psi^{i_{n}}(\eta_{n})\rangle^{\textrm{conn}}_{0}
\end{align}
the free energy is given by
\begin{equation}
\begin{aligned}
&F_{\textrm{QED}_{d}}=N_{f}F_{\rm free-ferm}(d)+F_{\rm Max.}^{(0)}(d)-\frac{1}{2}\log(e_0^2 R^{4-d})+
\frac{1}{2!}e_{0}^{2} G_{2} -\frac{1}{4!}e_{0}^{4} G_{4}+\ldots \\
&~~~~~~~~~~~+\Omega_{d}  R^{d-4}\big( d(d-1)(d-2)(d-3) b_{0}+d^2 c_{0}\big)\,,
\label{FQED}
\end{aligned}
\end{equation}
where $\Omega_{d}= 2\pi^{\frac{d+1}{2}}/\Gamma(\frac{d+1}{2})$ is the volume of the unit sphere, the free fermion free energy is given in (\ref{Ffer}),
and we have separated out the coupling dependent part $-\frac{1}{2}\log(e_0^2 R^{4-d})$ of the free Maxwell free energy, see eq.~(\ref{FMax}), (\ref{FMax0}). This
term plays an important role in the renormalization procedure upon using (\ref{e0Toe}), and its presence is necessary for cancellation of poles and to obtain a radius independent free energy at the fixed point.

The technical details of the calculation of $G_2$ and $G_4$ are given in Appendix \ref{AppG2G4}. To the order needed here, we find that their $\epsilon$ expansion is given by
\begin{align}
G_{2}=& N_{f}\Big(\frac{1}{6 \pi ^2 \epsilon }+\frac{4\big(5+3 (\log (4 \pi  R^2)+\gamma )\big)}{12^2 \pi ^2}+\frac{18 \pi ^2+124+  4 \big(5+3 (\log (4 \pi  R^2)+\gamma )\big)^2}{ 12^3 \pi ^2}\epsilon+\mathcal{O}(\epsilon^{2})\Big)\,, \notag\\
G_{4}=&\frac{N_{f}^2}{6 \pi ^4 \epsilon ^2}+\frac{N_{f} \big(8 N_{f} (5+3 (\log (4 \pi  R^2)+\gamma ))-18\big)}{12^2 \pi ^4 \epsilon }+\frac{1}{12^3 \pi ^4}\Big(16 N_{f}^2 \big(5+3 (\log (4 \pi  R^2)+\gamma )\big)^2\notag\\
&-72  N_{f}  \big(5+3 (\log (4 \pi  R^2)+\gamma )\big)+4 (77+9 \pi ^2)  N_{f}^2+9  N_{f}  (72 \zeta (3)-47)\Big)+ \mathcal{O}(\epsilon)\,.
\end{align}
Plugging these results into (\ref{FQED}), as well as the coupling renormalization (\ref{e0Toe}) and (\ref{abc-ren}), we find that all poles indeed cancel for arbitrary $e, b, c$. In particular, our calculation provides an independent check on the curvature counterterms (\ref{abc-ren}) to order $e^4$.

We can now compute the free energy at the IR fixed point by plugging in the critical couplings (\ref{estar}), (\ref{bc-star}). Defining
\begin{equation}
F_{\rm conf}(\epsilon) = F_{{\rm QED}_d}(e_*, b_*, c_*,\mu R)
\end{equation}
we find
\begin{align}
&F_{\rm conf}=N_{f}F_{\rm free-ferm}(d)+F_{\rm Max.}^{(0)}(d)+
\frac{1}{2} \log \Big(\frac{N_{f}}{6 \pi ^2\epsilon}\Big)\notag\\
&+ \Big(40+24 (\gamma+ \log (4 \pi ))+\frac{27}{N_{f}}\Big)\frac{\epsilon}{96}
+ \Big(\pi ^2+\frac{47}{9}
-\frac{9(8 N_f \zeta (3)-4 N_f+5)}{4N_f^2}\Big)\frac{\epsilon^2}{32}+\mathcal{O}(\epsilon^3)\,.
\label{dFIR}
\end{align}
Note that the result is indeed independent of the radius $R$, consistently with conformal invariance and the Callan-Symanzik equation (\ref{CS}).

Using the explicit $\epsilon$ expansion of the free Maxwell contribution, which can be obtained from (\ref{FMax0})
\begin{equation}
F_{\rm Max.}^{(0)}(d) =-\frac{1}{\sin\left(\frac{\pi d}{2}\right)} \left(\frac{31\pi}{90}+1.946\epsilon - 2.524 \epsilon^2 - 1.216 \epsilon^3+\mathcal{O}(\epsilon^4)\right)
\end{equation}
we then obtain our final result for $\tilde F_{\rm conf} = -\sin(\frac{\pi d}{2}) F_{\rm conf}$ given in eq.~(\ref{tF-conf}). As a test of this result, one can verify that to order $N_f^0$ it precisely agrees with the large $N_f$ prediction, eq.~(\ref{tF-largeN}) and (\ref{largeN-4mep}).

\section{Pad\'e approximation and the $F$-theorem}
\label{PadeSec}
\begin{figure}[h!]
                \centering
                \includegraphics[width=10cm]{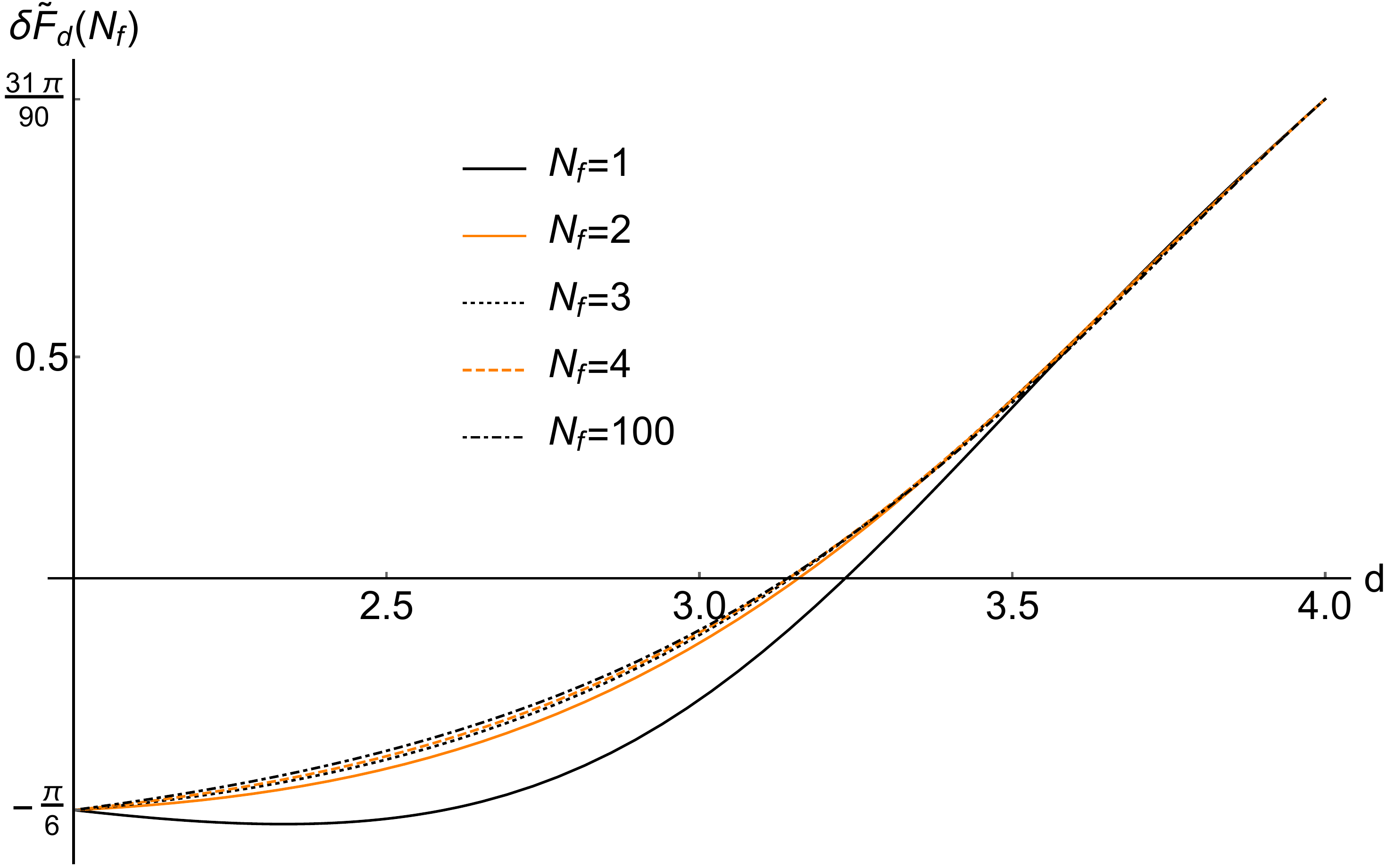}
                \caption{Pad\'e$_{[1,3]}$ on $\delta \tilde{F}_{d}(N_{f})$ for various $N_{f}$}
                \label{Pade1}
\end{figure}

\begin{figure}[h!]
                \centering
                \includegraphics[width=10cm]{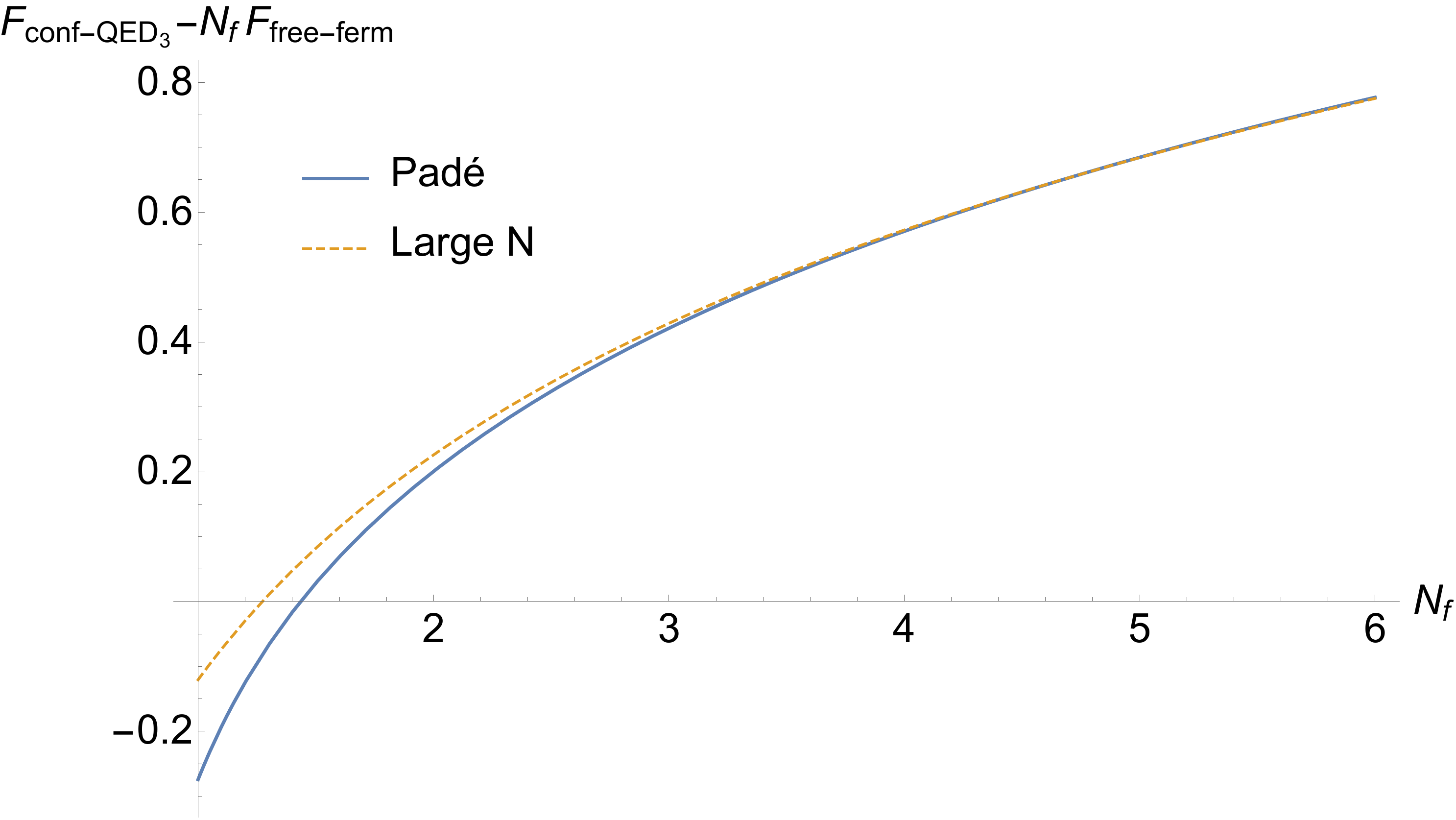}
                \caption{Comparison of the Pad\'e resummation of the $\epsilon$ expansion, and the large $N$ result (\ref{confQED}) for the sphere free energy of conformal QED$_3$.}
                \label{PadevsLN}
\end{figure}

A novel feature of the result (\ref{dFIR}) compared to the sphere free energy for the $O(N)$ Wilson-Fisher fixed points \cite{Fei:2015oha}, is the appearance of the $\log(\epsilon)$ behavior in $d=4-\epsilon$. This makes it difficult to apply standard resummation techniques like Pad\'e approximants. To circumvent this problem, we
isolate the logarithmic term $\frac{1}{2}\log(N_f/\epsilon)$ in $F$, which we treat exactly, and perform a Pad\'e extrapolation on the function
\begin{align}
\delta \tilde{F}_{d}(N_{f}) \equiv  \tilde{F}_{\rm conf}- N_{f}\tilde{F}_{\textrm{free-ferm}} +\frac{1}{2}\sin(\frac{\pi d}{2}) \log(\frac{N_{f}}{\epsilon})\, .
\end{align}
In $d=2$, the IR fixed point of QED$_d$ is the Schwinger model with $2 N_f$ massless two-component Dirac fermions. In the infrared it is described, via the non-abelian bosonization  \cite{Witten:1983ar}, by the level 1 $SU(2N_f)$ WZW model \cite{Gepner:1984au,Affleck:1985wa}. This is a CFT with $c=2N_f-1$
corresponding to $\tilde F = \frac{\pi}{6}(2N_f-1)$. Therefore, it is natural to use a two-sided Pad\'e approximant subject to the constraints:
\begin{align}
\delta \tilde{F}_{d}(N_{f}) =
\begin{cases}
-\frac{\pi}{6},\quad d=2\,, \\
\frac{31\pi}{90}-1.2597 \eps-0.6493 \eps^2+0.8429 \eps^3+\frac{0.4418 \eps^2}{N_{f}}-\frac{0.6203 \eps^3}{N_{f}}-\frac{0.5522 \eps^3}{N_{f}^2},\quad d=4-\epsilon\,.
\label{padeBC}
\end{cases}
\end{align}
This allows us to use Pad\'e approximants Pad\'e$_{[m,n]}$ of total order 4. The results for $d=3$ using these two-sided approximants Pad\'e$_{[2,2]}$ and Pad\'e$_{[1,3]}$ are given in table \ref{table1}. For comparison, we also present
the results using one-sided approximant  Pad\'e$_{[1,2]}$ obtained without assuming the boundary condition at $d=2$ (we see, however, that its agreement with the large $N_f$ expansion is not as good as that of both two-sided Pad\'e approximants).
\begin{table}[h]
\centering
\begin{tabular}{cccccccccc}
\hline
\multicolumn{1}{|c|}{$N_{f}$}         & \multicolumn{1}{c|}{1} & \multicolumn{1}{c|}{2} & \multicolumn{1}{c|}{3} & \multicolumn{1}{c|}{4} & \multicolumn{1}{c|}{5}& \multicolumn{1}{c|}{6}& \multicolumn{1}{c|}{10} & \multicolumn{1}{c|}{50}& \multicolumn{1}{c|}{100}\\ \hline
\multicolumn{1}{|c|}{Pad\'e$_{[2,2]}$}      & \multicolumn{1}{c|}{-} & \multicolumn{1}{c|}{-0.1512} & \multicolumn{1}{c|}{-0.1284} & \multicolumn{1}{c|}{-0.1237} & \multicolumn{1}{c|}{-0.1223}& \multicolumn{1}{c|}{-0.1218}& \multicolumn{1}{c|}{-0.1217}& \multicolumn{1}{c|}{-0.1230}& \multicolumn{1}{c|}{-0.1233}  \\ \hline
\multicolumn{1}{|c|}{Pad\'e$_{[1,3]}$} & \multicolumn{1}{c|}{-0.2743} & \multicolumn{1}{c|}{-0.1462} & \multicolumn{1}{c|}{-0.1284} & \multicolumn{1}{c|}{-0.1228} & \multicolumn{1}{c|}{-0.1204}& \multicolumn{1}{c|}{-0.1192} & \multicolumn{1}{c|}{-0.1176} & \multicolumn{1}{c|}{-0.1172}& \multicolumn{1}{c|}{-0.1172}\\ \hline
\multicolumn{1}{|c|}{Pad\'e$_{\textrm{average}}$} & \multicolumn{1}{c|}{-} & \multicolumn{1}{c|}{-0.1487} & \multicolumn{1}{c|}{-0.1284} & \multicolumn{1}{c|}{-0.1232} & \multicolumn{1}{c|}{-0.1213}& \multicolumn{1}{c|}{-0.1205} & \multicolumn{1}{c|}{-0.1196} & \multicolumn{1}{c|}{-0.1201}& \multicolumn{1}{c|}{-0.1203}\\ \hline
\multicolumn{1}{|c|}{Pad\'e$_{[1,2]}$} & \multicolumn{1}{c|}{-} & \multicolumn{1}{c|}{-0.1856} & \multicolumn{1}{c|}{-0.1259} & \multicolumn{1}{c|}{-0.1072} & \multicolumn{1}{c|}{-0.0986}& \multicolumn{1}{c|}{-0.0937} & \multicolumn{1}{c|}{-0.0861} & \multicolumn{1}{c|}{-0.0799}& \multicolumn{1}{c|}{-0.0793}\\ \hline
\multicolumn{1}{|c|}{$\eps$-expansion} & \multicolumn{1}{c|}{-0.7148} & \multicolumn{1}{c|}{-0.2113} & \multicolumn{1}{c|}{-0.1049} & \multicolumn{1}{c|}{-0.0632} & \multicolumn{1}{c|}{-0.0418}& \multicolumn{1}{c|}{-0.0291} & \multicolumn{1}{c|}{-0.0074}& \multicolumn{1}{c|}{0.0122}& \multicolumn{1}{c|}{0.0141} \\ \hline
\end{tabular}
\caption{Various Pad\'e approximations and the unresummed $\eps$-expansion of $\delta \tilde{F}_{d}(N_{f})$ at $d=3$. The two-sided approximants Pad\'e$_{[2,2]}$ and Pad\'e$_{[1,3]}$ are obtained
assuming the value $-\frac{\pi}{6}$ at $d=2$, while Pad\'e$_{[1,2]}$ does not use this assumption.
Row 3 is the average of the two-sided approximants, i.e. of the first two rows.
At large $N_{f}$ we expect to find $\delta \tilde{F}_{d=3}=\frac{1}{2}\log(\frac{\pi}{4}) \approx -0.1208 $.}
\label{table1}
\end{table}
We also plot the Pad\'e$_{[1,3]}$ approximant for different $N_{f}$ in figure \ref{Pade1}, as a function of $2<d<4$.
In figure \ref{PadevsLN}, we plot $F_{{\rm conf-QED}_3}-N_f F_{\rm free-ferm}$ comparing the result of the Pad\'e approximation and the large $N$ result (\ref{confQED}).

It is interesting to compare this result for the conformal phase of the theory with the $F$-value in the
broken symmetry phase. The latter contains $2N_f^2$ Goldstone bosons and a free Maxwell field,
which is dual to a scalar. \footnote{
In compact QED this scalar is compact
\cite{Polyakov:1975rs}, and it develops a vacuum expectation value.
If there are no monopole operators in the action, then the topological $U(1)_T$ symmetry is spontaneously broken 
and there is a massless scalar degree of freedom in the IR, just as in the 
non-compact case. We thank Z. Komargodski and S. Pufu for discussions on this.} 
At long distances, each of these fields is described by a conformally coupled
scalar field. Therefore, after the chiral symmetry breaking
\begin{align}
&F_{\textrm{SB}} (N_{f}) =   (2N_{f}^2+1) \Big(\frac{\log 2}{8} -\frac{3 \zeta(3)}{16 \pi^2} \Big)\, . 
\end{align}

\begin{figure}[h!]
\center{\includegraphics[width=0.5\linewidth]{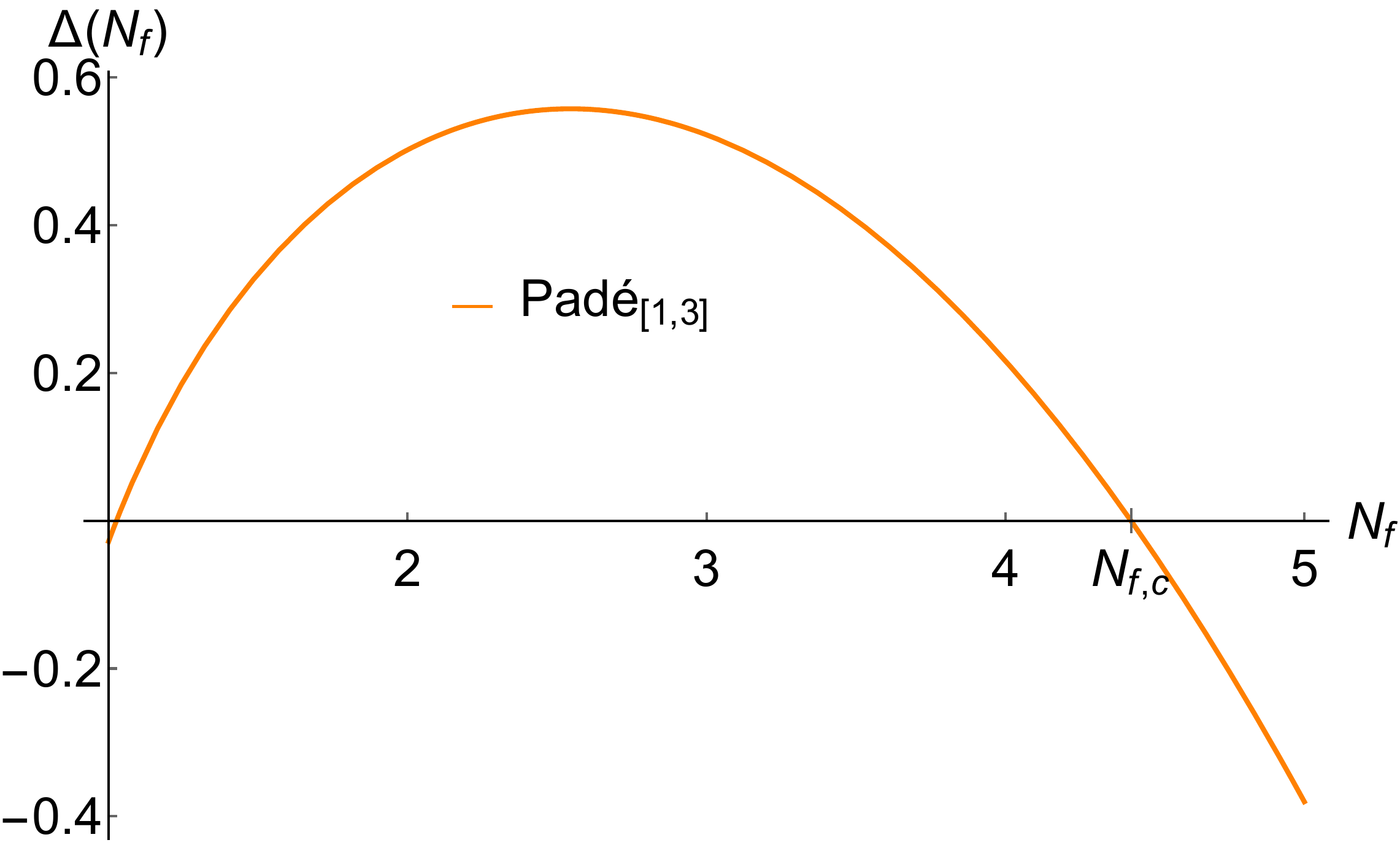}}
\caption{Plot of $\Delta (N_{f})=F_{{\rm conf}}(N_f)-F_{\rm SB}(N_f)$, using Pad\'e$_{[1,3]}$.}
\label{PadeDelta}
\end{figure}

To study if the $F$-theorem allows flow from the conformal phase to the phase with global
symmetry breaking,
we define the function
\begin{align}
&\Delta (N_{f}) =  F_{{\rm conf}} (N_{f}) - (2N_{f}^2+1) \Big(\frac{\log 2}{8} -\frac{3 \zeta(3)}{16 \pi^2} \Big)\, .
\end{align}
Its plot 
obtained using the Pad\'e$_{[1,3]}$ with $d=2$ boundary condition
is shown in Fig. \ref{PadeDelta}. One can also consider the corresponding function using the large $N_f$ expression (\ref{confQED}) for
$F_{\rm conf}$; this gives results that are close to those shown in Fig. \ref{PadeDelta}.

\begin{table}[h!]
\centering
\begin{tabular}{cccc}
\hline
\multicolumn{1}{|c|}{}         & \multicolumn{1}{c|}{Pad\'e$_{[1,3]}$} & \multicolumn{1}{c|}{Pad\'e$_{[2,2]}$} & \multicolumn{1}{c|}{Pad\'e$_{[1,2]}$}\\ \hline
\multicolumn{1}{|c|}{$N_{f,\textrm{c}}$ } & \multicolumn{1}{c|}{4.4204} & \multicolumn{1}{c|}{4.4180} & \multicolumn{1}{c|}{4.4530}  \\ \hline
\end{tabular}
\caption{Estimates of $N_{f,\textrm{c}}$, which is the solution of $F_{\rm conf}=F_{\rm SB}$, obtained from various Pad\'e approximants. Pad\'e$_{[1,3]}$ and Pad\'e$_{[2,2]}$ use the $d=2$ boundary condition in (\ref{padeBC}), while Pad\'e$_{[1,2]}$ only uses data from the $d=4-\epsilon$ expansion.}
\label{table1}
\end{table}

The plot in Fig. \ref{PadeDelta} implies that the RG flow from conformal to symmetry broken phase becomes impossible for $N_f$ between $4$ and $5$. This value of $N_f$ if found by solving the equation
$F_{\rm conf}= F_{\rm SB}$; it provides an upper bound on the integer value of $N_f$ where the
conformal window may become unstable: $N_{\rm crit}\leq 4$.


Now let us treat $N_f$ as a
continuous parameter and discuss the implications of conformal perturbation theory for the phase structure of the theory. As explained in the introduction, we expect that for $N_f$ near $N_{\rm crit}$, a $SU(2 N_f)$ invariant quartic operator is nearly marginal, and
we can work perturbatively in the small parameter $\delta=\Delta-3$.
The beta function for $\lambda$, the coefficient of the quartic operator, has the structure
$\beta_\lambda = \delta \lambda + A \lambda^2+ O(\lambda^3)$. Thus, in addition to the QED$_3$ fixed point at $\lambda=0$, we find a nearby fixed point at
$\lambda_*= -\delta/A$. For $N_f\gtrsim N_{\rm crit}$, this is a UV fixed point. It is another $SU(2N_f)$ invariant CFT which we could call QED$_3^*$. 
Its existence for $N_f$ slightly above $N_{\rm crit}$ is guaranteed by the conformal perturbation theory. It also exists for large $N_f$, where it is a double-trace deformation of QED$_3$. Therefore,
QED$_3^*$ may exist for all $N_f > N_{\rm crit}$.

\begin{figure}[h!]
                \centering
                \includegraphics[width=10cm]{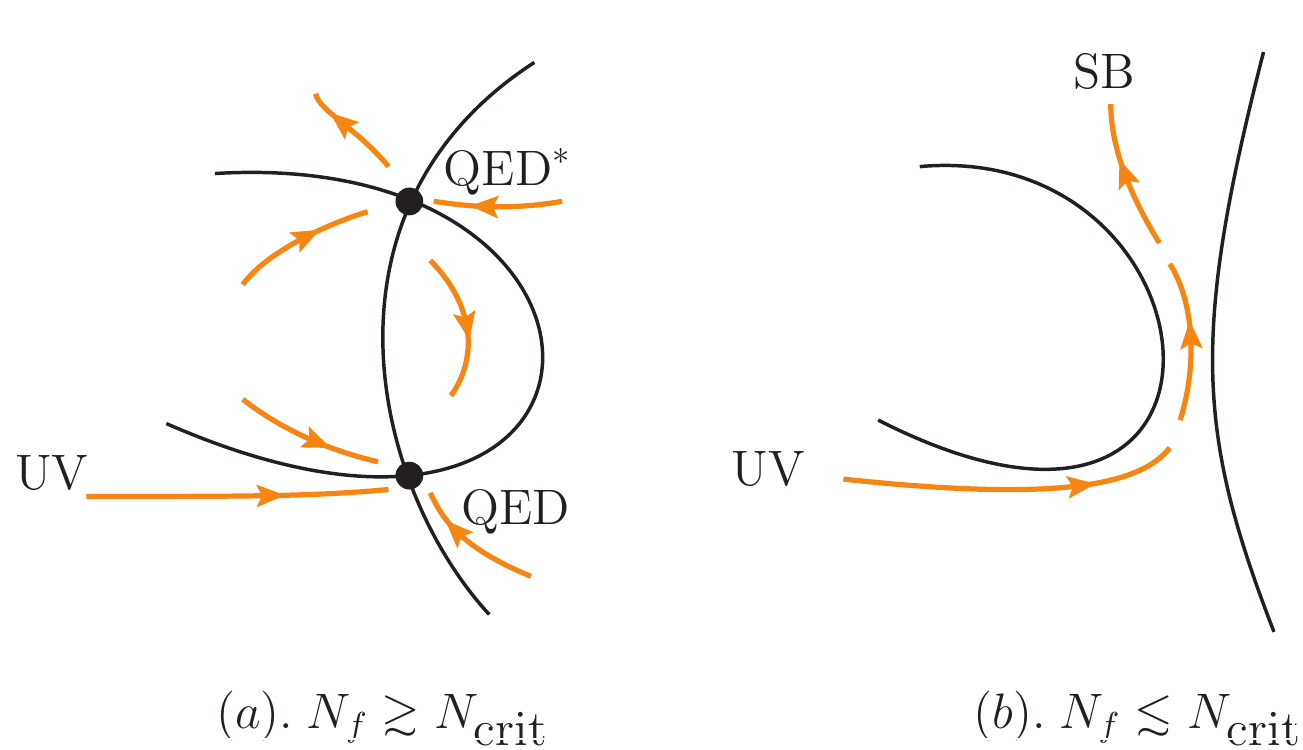}
                \caption{Schematic picture of RG flows for $N_f\gtrsim N_{\rm crit}$ (a) and $N_f\lesssim N_{\rm crit}$ (b).  The
QED$_3$ and QED$_3^*$ fixed points merge at $N_{\rm crit}$ and acquire small imaginary parts for 
$N_f\lesssim N_{\rm crit}$.  In the latter case, the interacting conformal behavior is no longer
possible, but the RG
flow from the UV can ``hover'' near the complex fixed points before running
away to large quartic coupling and presumably leading to the
broken symmetry phase. }
                \label{RGdiag}
\end{figure}

A commonly discussed scenario is that the
QED$_3$ and QED$_3^*$ fixed points merge at $N_{\rm crit}$ and acquire small imaginary parts for 
$N_f\lesssim N_{\rm crit}$ \cite{Kaveh:2004qa,Gies:2005as,Kaplan:2009kr,Braun:2014wja}.
This means that the interacting conformal behavior is
impossible for $N_f\lesssim N_{\rm crit}$, but the RG
flow from the UV can ``hover'' near the complex fixed points before running
away to large quartic coupling and presumably leading to the
broken symmetry phase (see figure \ref{RGdiag}). 
During the
hovering $F$ can be made parametrically close to $F_{\rm conf}(N_{\rm crit})$. This is why the $F$-theorem requires $F_{\rm conf}(N_{\rm crit})> F_{\rm SB}$ (a similar argument involving the continuity of $F$ was given in \cite{Grover:2012sp}). As we have seen, this gives a rather stringent bound
$N_{\rm crit} \lesssim 4.4$. 

An alternate possibility is that both fixed points stay real and go through each other. Then the QED$_3$ fixed point continues to exist even after the appearance of a relevant operator; this relevant operator may create flow from QED$_3$ to the broken symmetry phase. Therefore, the $F$-theorem bound on $N_{\rm crit}$ is the same as with the ``merger and annihilation'' scenario.

\section*{Acknowledgments}
We thank L. Di Pietro, T. Dumitrescu, L. Fei, Z. Komargodski, S. Pufu and V. Rychkov for useful discussions.
We are especially grateful to L. Di Pietro, Z. Komargodski, I. Shamir and E. Stamou for sharing a draft of their paper. I.R.K. also thanks the Weizmann Institute of Science for hospitality during the workshop ``Bootstrap 2015.''
The work of SG was supported in part by the US NSF under Grant No.~PHY-1318681.
The work of IRK and GT was supported in part by the US NSF under Grant No.~PHY-1314198.

\appendix

\section{Eigenvalues of the kernel  $K_{\mu\nu}$}
\label{AppK}
On the $S^{d}$  in stereographical coordinates the kernel $K_{\mu\nu}=-\langle J_{\mu} J_{\nu}\rangle $ has the form
\begin{align}
K_{\mu\nu}(x,y) = -C_{J}\frac{4R^{2}}{(1+x^{2})(1+y^{2})} \frac{\big(\delta_{\mu\nu}-2\frac{(x-y)_{\mu}(x-y)_{\nu}}{|x-y|^{2}}\big)}{s(x,y)^{2\Delta}}\,.\label{kern1}
\end{align}
We need to decompose the kernel on a sum of  vector spherical harmonics:
\begin{align}
K_{\mu\nu}(x,y)= \sum_{\ell,m} \sum_{s} \lambda_{\ell}^{(s)} Y_{\mu,\ell m}^{(s)*}(x)Y_{\nu, \ell m}^{(s)}(y)\,,
\end{align}
where $s$ denotes different types of vector spherical harmonics, $\ell$ is the principal angular quantum number and the range of $m$ for given $\ell$ and $s$  is $g_{\ell}^{(s)}$ (see (\ref{deg-1})-(\ref{deg-0})). The eigenvalue $ \lambda_{\ell}^{(s)}$ has no $m$ dependence because of rotational invariance.
Because of vector spherical harmonics are orthonormal to each other we have\footnote{We assume that $\int d^{d}x\sqrt{g_{x}} Y^{(s)*}_{\mu,\ell m}(x) Y^{\mu (s')}_{\ell' m'}(x)=\delta_{\ell \ell'}\delta_{mm'} \delta_{ss'}$.}
\begin{align}
 \lambda_{\ell}^{(s)}= \int d^{d}x d^{d}y\sqrt{g_{x}} \sqrt{g_{y}}  K^{\mu\nu}(x,y)Y_{\mu,\ell m}^{(s)}(x)Y_{\nu,\ell m}^{(s)*}(y)\,. \label{lam1}
\end{align}
We first consider   longitudinal vector harmonics \cite{Rubin:1984tc}
\begin{align}
Y_{\mu,\ell m}^{(0)}(x) = \frac{\nabla_{\mu} Y_{\ell m}(x)}{\sqrt{\ell(\ell+d-1)}}\,,
\end{align}
where $Y_{\ell m}(x)$ are usual scalar spherical harmonics\footnote{We assume that $\nabla^{2}Y_{\ell m} =-\ell(\ell+d-1)R^{-2}Y_{\ell m}$ and $\int d^{d}x \sqrt{g_{x}} Y_{\ell m}(x) Y^{*}_{\ell' m'}(x)= \delta_{\ell \ell'}\delta_{mm'}$.}. Integrating by parts in (\ref{lam1}) we get
\begin{align}
 \lambda_{\ell}^{(0)}= \frac{1}{\ell(\ell+d-1)}\int d^{d}x d^{d}y\sqrt{g_{x}} \sqrt{g_{y}}  \nabla_{\mu}\nabla_{\nu}K^{\mu\nu}(x,y)\frac{1}{g^{(0)}_{\ell}}\sum_{m}Y_{\ell m}(x)Y^{*}_{\ell m}(y)\,, \label{lam2}
\end{align}
where we sum over $m$ and divide by the degeneracy  $g^{(0)}_{\ell}$\;\footnote{We are allowed to do this sum because on the l.h.s $ \lambda_{\ell}^{(0)}$ doesn't depend on $m$.}.
We use that \footnote{Notice that the $Y_{\ell m}(x)$ harmonics have  a factor of $R^{-d/2}$, which is consistent with the formula (\ref{sumofY}) because  $\textrm{vol}(S^{d})= 2\pi^{\frac{d+1}{2}}R^{d}/\Gamma(\frac{d+1}{2})$ and $Y_{\mu,\ell m}^{(s)}$ have a factor of $R^{-d/2+1}$.}
\begin{align}
\frac{1}{g^{(0)}_{\ell}}\sum_{m}Y_{\ell m}(x)Y^{*}_{\ell m}(y) =  \frac{1}{\textrm{vol}(S^{d})} \frac{C_{\ell}^{(d-1)/2}(1- s^{2}(x,y)/2R^{2})}{C_{\ell}^{(d-1)/2}(1)}\,, \label{sumofY}
\end{align}
where $C^{(d-1)/2}_{\ell}(x)$ is the
Gegenbauer polynomial.
Now due to rotational invariance we may choose $y=0$ and  find\footnote{Notice that we can not just take $y=0$  in ($\ref{lam2}$) without having  the sum over $m$. }
\begin{align}
 \nabla_{\mu}\nabla_{\nu}K^{\mu\nu}(x,0) = -\frac{2C_J}{(4 R^{2})^{\Delta+1}}(\Delta-d+1)(2-d+2\Delta+ x^{2} d) \frac{(1+x^{2})^{\Delta}}{(x^{2})^{\Delta+1}}\,.
\end{align}
Therefore we obtain
\begin{align}
 \lambda_{\ell}^{(0)}=&-\frac{C_J (\Delta-d +1)} {2^{\Delta } R^{2\Delta -d}\ell (d+\ell-1)}\frac{\textrm{vol}(S^{d-1})}{C_{\ell}^{(d-1)/2}(1)} \notag\\
 &\times  \int_{-1}^{1} dz (1+z)^{\frac{d-2}{2}}(1-z)^{\frac{d}{2}-\Delta-2}(2+2 \Delta-d -(1-z) (1+\Delta -d))
 C_{\ell}^{(d-1)/2}(z)\,,
 \end{align}
where we have changed the variable $x^{2}= (1-z)/(1+z)$.  Calculating the integral we get\footnote{Here we used the integral
$\int_{-1}^{1} dz (1+z)^{\nu-\frac{1}{2}}(1-z)^{\beta}C_{\ell}^{\nu}(z) = (-1)^{\ell} \frac{2^{\beta+\nu+\frac{1}{2}}\Gamma(\beta+1)\Gamma(\nu+\frac{1}{2})\Gamma(2\nu+l)\Gamma(\beta-\nu+\frac{3}{2})}{\ell! \Gamma(2\nu)\Gamma(\beta-\nu-\ell +\frac{3}{2})\Gamma(\beta+\nu+\ell+\frac{3}{2})}$, which can be found with the help of \cite{GradRyz} 7.311.3 and the relation $C_{\ell}^{\nu}(z)=(-1)^{\ell}C_{\ell}^{\nu}(-z)$. Also we used that $C_{\ell}^{\frac{d-1}{2}}(1) = \frac{(\ell +d-2)!}{\ell!(d-2)!}$. }
\begin{align}
 \lambda_{\ell}^{(0)} =-C_{J}\frac{2^{d-2 \Delta }\pi ^{d/2} (\Delta -1)  \Gamma \left(\frac{d}{2}-\Delta \right)}{\Gamma (\Delta +1)}\frac{d-\Delta -1}{\Delta -1}\frac{\Gamma (l+\Delta )}{\Gamma (d+l-\Delta )}\frac{1}{R^{2\Delta-d}}\,.
 \end{align}

The eigenvalues $\lambda_{\ell}^{(1)}$ for transverse spherical harmonics can be easily found for $\Delta=d-1$, using the embedding formalism.
In this case the kernel has the form  \cite{Drummond:1977uy}
\begin{align}
K_{ab}(\eta_{1},\eta_{2}) = -C_{J} \frac{R^{-2}}{(2d-4)(d-1)} P_{ac} \frac{\delta_{cb}}{|\eta_{1}-\eta_{2}|^{2d-4}}\,,
 \end{align}
where the operator $ P_{ac}\equiv \frac{1}{2}L^2\delta_{ac}+L_{ad}L_{dc}-(d-1)L_{ac}$  acts on $\eta_{1}$, and $L_{ab}\equiv\eta_{a}\frac{\partial}{\partial \eta_b} -
\eta_{b}\frac{\partial}{\partial \eta_a}$ and $L^{2}\equiv L_{ab}L_{ab}$\,. Now using the decomposition \cite{Drummond:1977uy}
\begin{align}
\frac{\delta_{cb}}{|\eta_{1}-\eta_{2}|^{2d-4}} = \sum_{l,m}\sum_{s} (2R)^{4-d} \pi^{\frac{d}{2}} \frac{\Gamma(2-\frac{d}{2})\Gamma(\ell +d-2)}{\Gamma(d-2)\Gamma(\ell +2)} Y^{(s)}_{c,\ell m}(\eta_{1}) Y^{(s)}_{b,\ell m}(\eta_{2})\,
 \end{align}
and the property of the operator $P_{ab}$: $P_{ab}Y^{(s)}_{b,\ell m}=0$ for $s\neq 1$ and $P_{ab}Y^{(1)}_{b,\ell m}=-(\ell +1)(\ell+d-2)Y^{(1)}_{a,\ell m}$,
we find
\begin{align}
K_{ab}(\eta_{1},\eta_{2}) = \sum_{\ell, m} C_{J} \frac{(2R)^{4-d}\pi^{\frac{d}{2}}(\ell+1)(\ell+d-2)\Gamma(2-\frac{d}{2})\Gamma(\ell+d-2)}{R^{2}(2d-4)(d-1)\Gamma(d-2)\Gamma(\ell+2)}
Y^{(1)}_{a,\ell m}(\eta_{1}) Y^{(1)}_{b,\ell m}(\eta_{2})\,,
 \end{align}
therefore for $\lambda_{\ell}^{(1)}$ we get
\begin{align}
\lambda_{\ell}^{(1)} =C_J \frac{ \pi ^{\frac{d}{2}}  \Gamma \left(2-\frac{d}{2}\right) \Gamma (\ell+d-1)}{2^{d-3} \Gamma (d)\Gamma (\ell+1)} \frac{1}{R^{d-2}}\,,
 \end{align}
which coincides with (\ref{lambs}) for $\Delta=d-1$.

\section{Calculation of $G_{2}$ and $G_4$}
\label{AppG2G4}

Using the propagators and vertex in (\ref{proppsi}),(\ref{propA}), (\ref{vertex})  we find for the two-point function
\begin{align}
G_{2} = -N_{f}\int d^{d}\eta_{1}d^{d}\eta_{2}\, \tr (Q_{ab}(\eta_{1})\alpha_{b} S(\eta_{1},\eta_{2}) Q_{cd}(\eta_2) \alpha_{d} S(\eta_{2},\eta_{1})) D_{ac}(\eta_{1},\eta_{2})\,.
\end{align}
After calculation we obtain
\begin{align}
G_{2} &= -\frac{(N_{f}\tr {\bf 1} )\Gamma(\frac{d}{2})\Gamma(d-2)}{2^{d+2}\pi^{\frac{3d}{2}} R^{d-2}}\int d^{d}\eta_{1}d^{d}\eta_{2} \frac{\big(d-2+\frac{1}{2R^{2}}(\eta_{1}-\eta_{2})^{2}\big)}{|\eta_{1}-\eta_{2}|^{2d-2}}\,_{2}F_{1}(1,d-2,\frac{d}{2},1-\frac{(\eta_{1}-\eta_{2})^{2}}{4R^{2}})\,.
\end{align}
Due to rotational invariance we can put $\eta_{2}$ to the north pole of the sphere and get in stereographic coordinates
\begin{align}
G_{2}
&=-\frac{(N_{f}\tr {\bf {\bf 1}})R^{4-d} \Gamma (d-2)}{(16 \pi )^{\frac{d-1}{2}} \Gamma \left(\frac{d+1}{2}\right)}\int_{0}^{\infty} \frac{dx}{(1+x^{2})x^{d-1}}\Big(d-\frac{2}{1+x^{2}}\Big)\,_{2}F_{1}(1,d-2,\frac{d}{2},\frac{1}{(1+x^{2})})\,.
\end{align}
Now introducing the variable $z= 1/(1+x^{2})$ we find
\begin{align}
G_{2} &=-\frac{(N_{f}\tr {\bf 1})R^{4-d} \Gamma (d-2)}{ 2(16\pi) ^{\frac{d-1}{2}} \Gamma \left(\frac{d+1}{2}\right)}
\int_{0}^{1} dz z^{\frac{d}{2}-1}(1-z)^{-\frac{d}{2}}(d-2z)\,_{2}F_{1}(1,d-2,\frac{d}{2},z)\,.
\end{align}
The integral can be calculated exactly and we obtain\footnote{$G_{2}$ was also computed in \cite{Shore:1978hj, Harris:1993gd } by different methods. }
\begin{align}
G_{2} &=-\frac{(N_{f}\tr {\bf 1}) R^{4-d} \Gamma (d-1)}{(d-3) 4^{\frac{d+2}{2}} (4 \pi )^{\frac{d-3}{2}} \sin \left(\frac{\pi  d}{2}\right) \Gamma \left(\frac{d+1}{2}\right)}\,.
\end{align}

\begin{figure}[h!]
                \centering
                \includegraphics[width=4cm]{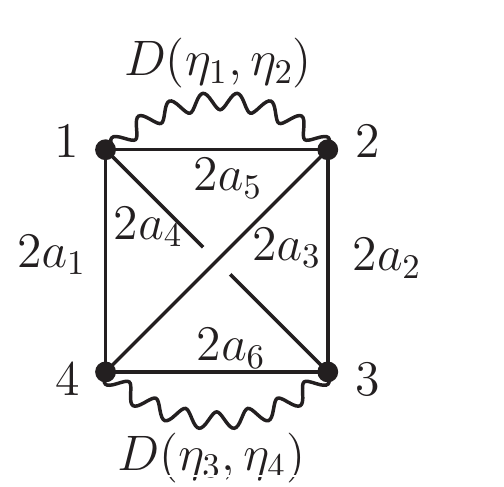}
                \caption{The diagrammatic representation for the  integral $I_{4}(a_{1},...,a_{6})$.}
                \label{I4gener}
\end{figure}

After doing combinatorics  for $G_{4}$ we find that it consists of the sum of $34$ integrals  of the form
\begin{align}
&I_{4}(a_{1},...,a_{6}) =\notag\\
&=\Big(\frac{\Gamma(\frac{d}{2})}{2\pi^{\frac{d}{2}}}\Big)^{4}\Big(\frac{R^{2-d}\Gamma(d-2)}{(4\pi)^{\frac{d}{2}}\Gamma(\frac{d}{2})}\Big)^{2}\int \prod_{i=1}^{4}d^{d}\eta_{i}\frac{\;_{2} F_{1}(1,d-2,\frac{d}{2}; 1-\frac{s(\eta_{1},\eta_{2})^{2}}{4R^{2}})\,_{2} F_{1}(1,d-2,\frac{d}{2}; 1-\frac{s(\eta_{3},\eta_{4})^{2}}{4R^{2}})}{s(\eta_{1},\eta_{2})^{2a_{1}}s(\eta_{2},\eta_{3})^{2a_{2}}s(\eta_{3},\eta_{4})^{2a_{3}}s(\eta_{1},\eta_{4})^{2a_{4}}s(\eta_{1},\eta_{3})^{2a_{5}}s(\eta_{2},\eta_{4})^{2a_{6}}},
\end{align}
where we used the exact form of the photon propagator (\ref{propA}).
The integral $I_{4}$ is represented diagrammatically in figure \ref{I4gener}.

\noindent The next step is to write the integral  $I_{4}$ in the Mellin-Barnes (MB) representation and then use the Mathematica program \cite{Czakon:2005rk, Smirnov:2009up} to find $I_{4}$ as a series in $\epsilon$ \cite{Smirnov:2012gma}. For the  photon propagator we use the Mellin-Barnes representation\footnote{We use the formula $\,_{2}F_{1}(a,b,c;x) = \frac{\Gamma(c)}{\Gamma(a)\Gamma(b)\Gamma(c-a)\Gamma(c-b)} \frac{1}{2\pi i}\int_{-i\infty}^{+i\infty} dz \Gamma(-z) \Gamma(a+z)\Gamma(b+z)\Gamma(c-a-b-z)(1-x)^{z}$.}
\begin{align}
D(\eta_{1},\eta_{2})=-\frac{\sin \left(\frac{\pi  d}{2}\right)}{2^d \pi ^{\frac{d}{2}+1} R^{d-2}} \frac{1}{2\pi i}\int_{-i\infty}^{+i\infty}  dz \Gamma(-z) \Gamma(1+z)\Gamma(d-2+z)\Gamma(1-\frac{d}{2}-z)\Big(\frac{s(\eta_{1},\eta_{2})^{2}}{4R^{2}}\Big)^{z}\,.
\end{align}
Then using  methods similar as discussed in \cite{Fei:2015oha} one can write the general MB form for the integral
\begin{align}
I_{4}(a_{1},...,a_{6})=&\frac{ \sin ^2(\frac{\pi  d}{2}) \Gamma (\frac{d}{2})^4}{2^{d+7} \pi ^{\frac{1}{2} (5 d+3)} \Gamma (\frac{d+1}{2})}(2 R)^{2 d+4-2\sum_{i=1}^{6}a_{i}}\frac{1}{(2\pi i)^{5}}\int \prod_{i=1}^{5}dz_{i} \Gamma (-z_{1}) \Gamma (-z_{2})\Gamma (z_{1}+1)  \Gamma (z_{2}+1) \notag\\
&\Gamma (d-2+z_{1}) \Gamma (1-\frac{d}{2}-z_{1}) \Gamma (d-2+z_{2}) \Gamma (1-\frac{d}{2}-z_{2})\Gamma_{2}(d-a_{145}+z_{1},a_{1},a_{5}-z_{1}|z_{3},z_{4},z_{5})\notag\\
&\Gamma_{0}(d-a_{235}+z_{1}-z_{4},d-a_{136}+z_{2}-z_{3},a_{3}-z_{5})\,, \label{MBI4}
\end{align}
where  $a_{mnk...}\equiv a_{m}+a_{n}+a_{k}+...$ and $\Gamma$-blocks are
\begin{align}
\Gamma_{0}(a_{1},a_{2},b)&=\frac{\pi^{d}}{\Gamma(\frac{d}{2})} \frac{\Gamma(\frac{d}{2}-b)\Gamma(a_{1}+b-\frac{d}{2})\Gamma(a_{2}+b-\frac{d}{2})\Gamma(a_{1}+a_{2}+b-d)}{\Gamma(a_{1})\Gamma(a_{2})\Gamma(a_{1}+a_{2}+2b-d)}\,, \notag\\
\Gamma_{2}(a,b_{1},b_{2}|z_{1},z_{2},z_{3})&=\frac{\pi^{d/2} \prod\limits_{i=1}^{3}\Gamma(-z_{i})\Gamma(a+b_{1}+b_{2}-\frac{d}{2}+\sum_{i=1}^{3}z_{i}) }{\Gamma(a)\Gamma(b_{1})\Gamma(b_{2})\Gamma(d-a-b_{1}-b_{2})} \Gamma(b_{1}+z_{1}+z_{3})\Gamma(b_{2}+z_{2}+z_{3})
  \notag\\
&\qquad\times  \Gamma(d-a-2b_{1}-2b_{2}-z_{1}-z_{2}-2z_{3})\,.
\end{align}
For some values of $a_{1},...,a_{6}$ the MB representation (\ref{MBI4}) is divergent or exactly zero for any $d$ (due to the term $\Gamma(0)$). To handle this problem we used an additional regulator $\delta$, so we were consistently calculating the integrals
\begin{align}
I^{\textrm{reg}}_{4}(a_{1},...,a_{6})=I_{4}(a_{1}+\delta,...,a_{6}+\delta)\,.
\end{align}
Each integral out of $34$ in $G_{4}$ depends on $\delta$ and $\epsilon$. But the  $G_{4}$ itself is free of $\delta$ and depends only on $\epsilon$ as it should.

In general MB approach gives a  result in terms of  a series in $\epsilon$ and $\delta$ and each coefficient of this series is a sum of convergent Melling-Barnes integrals, which are independent of $\epsilon$ and $\delta$ and  can be calculated numerically.  The numerical result is usually equal to some exact combinations of constants like $\pi^{4}, \zeta(3)$, e.t.c.
To get an exact answer we used integer relation search algorithm PSLQ  \cite{BaileyDH}. The final result reads
\begin{align}
G_{4}=&\frac{N_{f}^2}{6 \pi ^4 \epsilon ^2}+\frac{N_{f} \big(8 N_{f} (5+3 (\log (4 \pi  R^2)+\gamma ))-18\big)}{12^2 \pi ^4 \epsilon }+\frac{1}{12^3 \pi ^4}\Big(16 N_{f}^2 \big(5+3 (\log (4 \pi  R^2)+\gamma )\big)^2\notag\\
&-72  N_{f}  \big(5+3 (\log (4 \pi  R^2)+\gamma )\big)+4 (77+9 \pi ^2)  N_{f} ^2+9  N_{f}  (72 \zeta (3)-47)\Big)\,.
\end{align}
Note that in this calculation of $G_{4}$ we have used $d=4-\epsilon$ and $N= N_{f}\tr {\bf 1}= 4N_{f}$.

\bibliographystyle{ssg}
\bibliography{F-QED}

\end{document}